\documentclass{aa}
\usepackage{graphicx}   

\def\void#1{{}}

\begin{document}

\title{The SWIRE-VVDS-CFHTLS surveys: stellar mass assembly over
 the last 10 Gyrs. Evidence for a major build up of the red sequence
 between z=2 and z=1.
\thanks{Based on data obtained with the European southern observatory Very Large Telescope, Paranal, Chile, program 070A-9007(A) and on observations obtained with MegaPrime/MegaCam, a joint project of CFHT and CEA/DAPNIA, at the Canada-France-Hawaii Telescope (CFHT) which is operated by the National Research Council (NRC) of Canada, the Institut National des Science de l'Univers of the Centre National de la Recherche Scientifique (CNRS) of France, and the University of Hawaii. This work is based in part on data products produced at TERAPIX and the Canadian Astronomy Data Centre as part of the Canada-France-Hawaii Telescope Legacy Survey, a collaborative project of NRC and CNRS and on data obtained as part of the UKIRT Infrared Deep Sky Survey
  }}

\author{ 
 S. Arnouts\inst{1,2}
\and C.J. Walcher      \inst{2}
\and O. Le F\`evre     \inst{2}
\and G. Zamorani      \inst{3} 
\and O. Ilbert             \inst{4}
\and V. Le Brun         \inst{2}
\and L. Pozzetti         \inst{3} 
 \and S. Bardelli         \inst{3}
\and L. Tresse           \inst{2}
\and E. Zucca            \inst{3}
\and S. Charlot           \inst{5}
\and F. Lamareille      \inst{6}
\and H.J. McCracken \inst{5}
\and M. Bolzonella      \inst{3} 
\and A. Iovino             \inst{7}
\and C. Lonsdale \inst{8}
\and M. Polletta   \inst{8}
\and J. Surace     \inst{9}
\and D. Bottini \inst{10}
\and B. Garilli \inst{10}
\and D. Maccagni \inst{10}
\and J.P. Picat \inst{6}
\and R. Scaramella \inst{3}
\and M. Scodeggio \inst{10}
\and G. Vettolani \inst{3}
\and A. Zanichelli \inst{3}
\and C. Adami    \inst{2}
\and A. Cappi     \inst{3}
\and P. Ciliegi    \inst{3}  
\and T. Contini  \inst{6}
\and S. de la Torre \inst{2}
\and S. Foucaud \inst{14}
\and P. Franzetti \inst{10}
\and I. Gavignaud \inst{15}
\and L. Guzzo \inst{7}
\and B. Marano     \inst{3}  
\and C. Marinoni \inst{13}
\and A. Mazure \inst{2}
\and B. Meneux \inst{7}
\and R. Merighi   \inst{3} 
\and S. Paltani \inst{12}
\and R. Pell\`o \inst{6}
\and A. Pollo \inst{2}
\and M. Radovich \inst{11}
\and S. Temporin \inst{7}
\and D. Vergani \inst{10}
	}
\offprints{St\'ephane Arnouts,
  \email{arnouts@cfht.hawaii.edu}
	  }
\institute{ 
 Canada-France-Hawaii Telescope Corporation, Kamuela, HI-96743, USA 
    \and
 Laboratoire d'Astrophysique de Marseille, UMR 6110, BP8, 13376 Marseille Cedex 12, France
\and
 INAF-Osservatorio Astronomico di Bologna - Via Ranzani,1, 40127, Bologna, Italy
\and
 Institute for Astronomy, 2680 Woodlawn Dr., University of Hawaii,Honolulu, Hawaii, 96822
\and
Institut d'Astrophysique de Paris, UMR 7095, 98 bis Bvd Arago, 75014 Paris, France
\and
 Laboratoire d'Astrophysique de Toulouse, UMR 5572,14  av. E. Belin, 31400 Toulouse, France
\and
INAF-Osservatorio Astronomico di Brera - Via Brera 28, Milan, Italy
\and
University of California, San Diego 9500 Gilman Dr. La Jolla, CA
92093-0424, USA
\and 
Spitzer Science Center, 
Mail Stop 314-6, 1200 East California Boulevard, Pasadena, CA 91125, USA  
\and
 IASF-INAF - via Bassini 15, I-20133, Milano, Italy
\and
 INAF-Osservatorio Astronomico di Capodimonte - Via Moiariello 16, I-80131, Napoli, Italy
\and
Geneva Observatory, ch. des Maillettes 51, CH-1290 Sauverny, Switzerland
\and 
Centre de Physique Th\'eorique, UMR 6207 CNRS-Universit\'e de Provence, F-13288 Marseille France
\and	
School of Physics \& Astronomy, University of Nottingham, University Park, Nottingham, NG72RD, UK      
\and 
 Astrophysical Institute Potsdam, An der Sternwarte 16, 14482  Potsdam, Germany
	  }
\date{Received : 12 April 2007 ,  Accepted : 18 September 2007 }
\titlerunning{K-LFs and stellar mass density up to z=2}
\authorrunning{Arnouts et al. }

\abstract{ We present  an analysis of the stellar mass growth over  
 the last 10 Gyrs ($z\le 2$) using a unique large sample of galaxies selected
 at  $3.6\mu m$. We have assembled accurate 
 photometric  and spectroscopic  redshifts for $\sim$ 21200 and 1500
 galaxies, respectively, with F(3.6$\mu m)\ge 9.0\mu$Jy  by combining
 data  from Spitzer-SWIRE IRAC, the VIMOS VLT Deep Survey (VVDS), 
 UKIDSS  and very deep optical CFHTLS  photometry.
 We split our sample into quiescent (red) and active (blue) galaxies
 on the basis of an SED fitting procedure that we have compared  with 
 the strong rest-frame color  bimodality $(NUV-r')_{ABS}$. 
 The present sample contains  $\sim$ 4400 quiescent galaxies. 
 Our measurements of the K-rest frame luminosity function and luminosity density 
 evolution support the idea that a large fraction of galaxies is already assembled
 at $z\sim 1.2$, with almost 80\% and 50\% of the active and quiescent populations
 already in place, respectively. \\
 Based on the analysis of the evolution of the stellar mass-to-light ratio (in K-band)
 for the spectroscopic sub-sample, we derive the stellar mass density for the 
 entire sample. We find that the global evolution of the stellar mass density
 is well reproduced by the star formation rate derived from UV based measurements 
 when an appropriate dust correction is applied, which supports the idea 
 of an initial mass function that is on average universal.\\
  Over the last 8Gyrs ($z\le 1.2$) we observe that the stellar mass density 
 of the active population shows  a modest mass growth rate 
 ($\dot{\rho} \sim0.005(\pm0.005)\  M_{\odot}/Mpc^3/yr$), consistent with a constant 
 stellar mass density,  $\rho_{\star}^{active}\sim 3.1\  10^{8} M_{\odot}/Mpc^3$.
 In contrast,  an increase by a factor of  $\sim 2$ for the quiescent population over
 the same timescale is observed. 
 As a consequence, the growth of the stellar mass in the quiescent population must
 be due to the shutoff of star formation in active galaxies that migrate into the
 quiescent population. We estimate this  stellar mass flux  to be 
 $\dot{\rho}_{A\rightarrow Q} \sim0.017(\pm0.004) M_{\odot}/Mpc^3/yr$, which 
 balances the major fraction of new stars born according to  
 our best SFR estimate ($\dot{\rho}=0.025(\pm0.003) M_{\odot}/Mpc^3/yr$).\\
 From $z=2$ to $z=1.2$,  we observe a major build-up of the quiescent 
 population  with an  increase by a factor of $\sim$10 in stellar mass 
 (a mass growth rate of  $\sim0.063 M_{\odot}/Mpc^3/yr$).
 This  rapid evolution
  suggests that we are observing the epoch when, for the first time in the
 history of the universe, an increasing fraction of galaxies end their star
 formation activity and start to build up the red sequence.
}
\maketitle
\section{Introduction}

 The  strong decline of the cosmic star formation rate (SFR) since $z\sim 1$ 
 is now well established  (Schiminovich et al.,  2005; Hopkins et al., 2006),
 and has been shown to be accompanied by a decrease of faint star forming 
 galaxies and the decline of luminous ultra-violet galaxies 
 (Lilly et al., 1996; Arnouts et al., 2005).
 One fundamental question is to understand the link between 
 the global decline of the star formation rate and the evolution of the mass 
 assembly.   Inspection of the specific star formation rate 
 (SFR per unit of stellar mass)  in individual galaxies reveals that the preferred 
 site of star formation activity has migrated  from massive systems at high z 
 to low mass systems at low-z. This is usually referred to as the downsizing effect
 (Cowie et al., 1996; Juneau et al., 2005). Evidences of such an effect
 have been seen in the fundamental plane relation of elliptical
 galaxies (Treu et al., 2005) as well as in the analysis of spectra of
 local galaxies  (Heavens et al., 2004; Kauffmann et al., 2003) where
 both analyses show that massive galaxies have formed their stars earlier
 than less massive ones. 
 Clustering properties of star-forming galaxies at high and low redshifts 
 reveal  that the bulk of star formation activity has migrated from massive
 dark matter halos (DMH) at  high z to low mass DMH at low z, generalizing
 the downsizing effect from  stellar mass to the dark matter (Heinis et al., 2007).

 While the $\Lambda$CDM hierarchical scenario 
 successfully describes the  clustering properties of galaxies on large scales  
 (Mo and White, 1996 ;  Springel et al., 2006), the quenching of star formation 
 in massive systems  is less well understood. It relies on the complex physics of
 baryons and there is not yet enough constraints on the mechanisms involved
 in  the regulation of star formation. 
 These mechanisms may act on galactic scales like
 AGN, Supernovae feedback (Benson et al., 2003,  Croton et al., 2006) or
 on large scales via merging or gas heating in dense environments
 (Naganime et al., 2001; Yoshikawa et al., 2001).  
 For instance, to reproduce the properties of local massive elliptical galaxies, 
 de Lucia et al. (2006) show that the stars could form first at high redshift
 in sub-galactic units, while the galaxies would continue their stellar mass assembly
 through merging onto low redshift.

 In this context, the most stringent constraint for the semi-analytical models
 is the direct observation of the number density of massive galaxies at high
 redshift.
 Evidences of the existence of massive galaxies at high redshift 
 are becoming numerous. An important population
 of EROs at $z\ge 1$ has been discovered 
 (K20 survey: Cimatti et al., 2002), with roughly half of them 
 being old stellar systems (Cimatti et al., 2004). They could be the already  
 assembled progenitors of local massive ellipticals, as supported by
 their clustering signals and number density  (Daddi et al., 2002;  Arnouts, 2003). 
 The measures of the evolution of the stellar mass density from 
 NIR surveys  found  that half of the stellar mass is already
 in place at $z\le 1-1.5$ (Fontana et al., 2003; 2004;  Pozzetti et al., 2003;
 Drory et al., 2004; 2005;  Caputi et al., 2005) moving the formation 
 epoch close to the peak of the cosmic SFR.
 The analysis of the galaxy mass function (GMF) at different redshifts provides 
 additional information on how the process of mass assembly acts on 
 different mass scales. Such analysis is becoming common 
 among the recent surveys (Fontana et al., 2004, 2006;  Drory et al., 2004, 2005;  
 Bundy et al., 2005;  Borch et al., 2006; Franceschini et al., 2006, Pozzetti et al.,
 2007).  While the different studies do not necessarily agree with each other, 
 it emerges that the most massive galaxies have undergone less evolution
 than less massive ones since  $z\sim 1.0$  and  50 to 70\% are already in place 
 at such a redshift, revealing a faster assembly for the most massive systems.

 An alternative approach to measure the density of massive galaxies at high
 redshift is to investigate the evolution of the  local elliptical galaxies that dominate 
 the  massive end of the GMF (Bell et al., 2003; Baldry et al., 2004, Bundy et al., 2006, 
 Cirasuolo et al., 2006).
 This has been done by  following the redshift evolution of the red sequence, defined 
 by the $(U-V)$ or $(U-B)$ colors, from today up to $z\sim 1.1$. Bell et al. (2004) 
 and Faber et al. (2005)  find that the density of red galaxies drops  by a factor 4 or 2.5 
 respectively, suggesting that their stellar mass density roughly doubles, while at the same
 time the blue population shows little evolution. They interpret this evolution by the 
 migration of blue galaxies that quenched their star formation and migrated into the
 red sequence. 
 Bell et al. (2004) also reported a significant evolution of the most massive ellipticals.
 Because there is a shortage of massive blue galaxies able to produce
 those massive ellipticals, they introduce the idea of "dry" or purely stellar mergers
 (merging between red galaxies) as a possible scenario to produce these massive 
  local ellipticals.

  Analysis from the VVDS  survey by Zucca et al. (2006), who define the early-type 
  class based on SED fitting, 
 report  a  more modest decline ($\sim$40\%) in the number density of this population
 up to $z\sim1.1$,  
 which is  also consistent with the analysis by Ilbert et al. (2006) based on a 
 morphologically selected elliptical sample in the CDF  South.\\
 Similarly to VVDS results, Brown et al. (2007), who have selected ellipticals
 with  same optical color criteria than Bell et al. (2004),  in an area of 7 deg$^2$,    
 do not observe any  evolution of  comoving density ($\Phi^{\star}$)  up
 to $z=1$ and according to the modest luminosity density increase ($\sim$36\%),
  predicts  that  the stellar mass has roughly doubled since  $z\sim 1$.
 Moreover, they observe a modest evolution of luminous ellipticals
 (with $L\ge 4L^{\star}$) with 80\% of their stellar mass in place at $z\sim 0.8$, 
 suggesting that "dry" mergers should not play a dominant role in the evolution
 of ellipticals over the last 8Gyrs.

 Complementary analysis by Bundy et al. (2006) find signatures of 
 a downsizing effect  with quenching affecting first the massive
 galaxies and then moving to lower mass systems. 
 This can be interpreted as an anti-hierarchical  process with massive
 early types ending first their mass assembly  while the less massive ones
 are still assembling their mass  (see also Cimatti et al., 2006). 
 If such a process is confirmed, it  would be a  strong observational
 constraint  for the models of galaxy formation within the standard
 paradigm.    

 A large fraction of the surveys discussed above are  limited to $z\le 1-1.2$, 
  due to optical selection, while the deepest surveys are still covering 
 modest sizes in particular when the sample is splitted by galaxy types. 
 The impact of cosmic variance on results coming from small fields remains  
 a major problem. Even in the three fields covered by COMBO17, it still 
 appears to  play a significant role (Bell et al., 2004; Somerville et al., 2004).
 The selection of  the samples varies from optical to IR band
 and the  stellar mass is estimated from various approaches including SED
 fitting, rest-frame colors, with or without infrared information, introducing 
 significant dispersions in the final estimates. 
 Additionally, the selection of  elliptical galaxies based on optical colors has
  been found to show  a significant contamination by Sey2 AGN and star forming
  galaxies  up to 40\%  (Franzetti et al., 2007).
 
 In this paper we explore the redshift evolution of the stellar mass assembly for 
 a unique large sample selected in the observed Mid-Infrared, via the analysis of the
 K-rest luminosity functions, the luminosity density  and  the evolution of the 
 mass to light ratio.  We take advantage of a native IRAC selection that allows
 us to detect massive red objects  up to $z\sim 2$, and we isolate a large sample 
 of red/quiescent galaxies according to their low level of star formation  activity.
 We then discuss implications for the stellar mass assembly of the active and
 quiescent  populations.
 Throughout the paper we adopt the concordance cosmology 
 $\Omega_m=0.3$ and $\Omega_{\Lambda}=0.7$ and
 $H_0=70 km.s^{-1} Mpc^{-1}$. All magnitudes are in the AB system, 
 and for convenience we adopt a Salpeter IMF unless otherwise specified.
\label{sec:intro}
%
\section{The data }
\label{sec:data}
%
\begin{figure}
  \resizebox{\hsize}{!}{\includegraphics{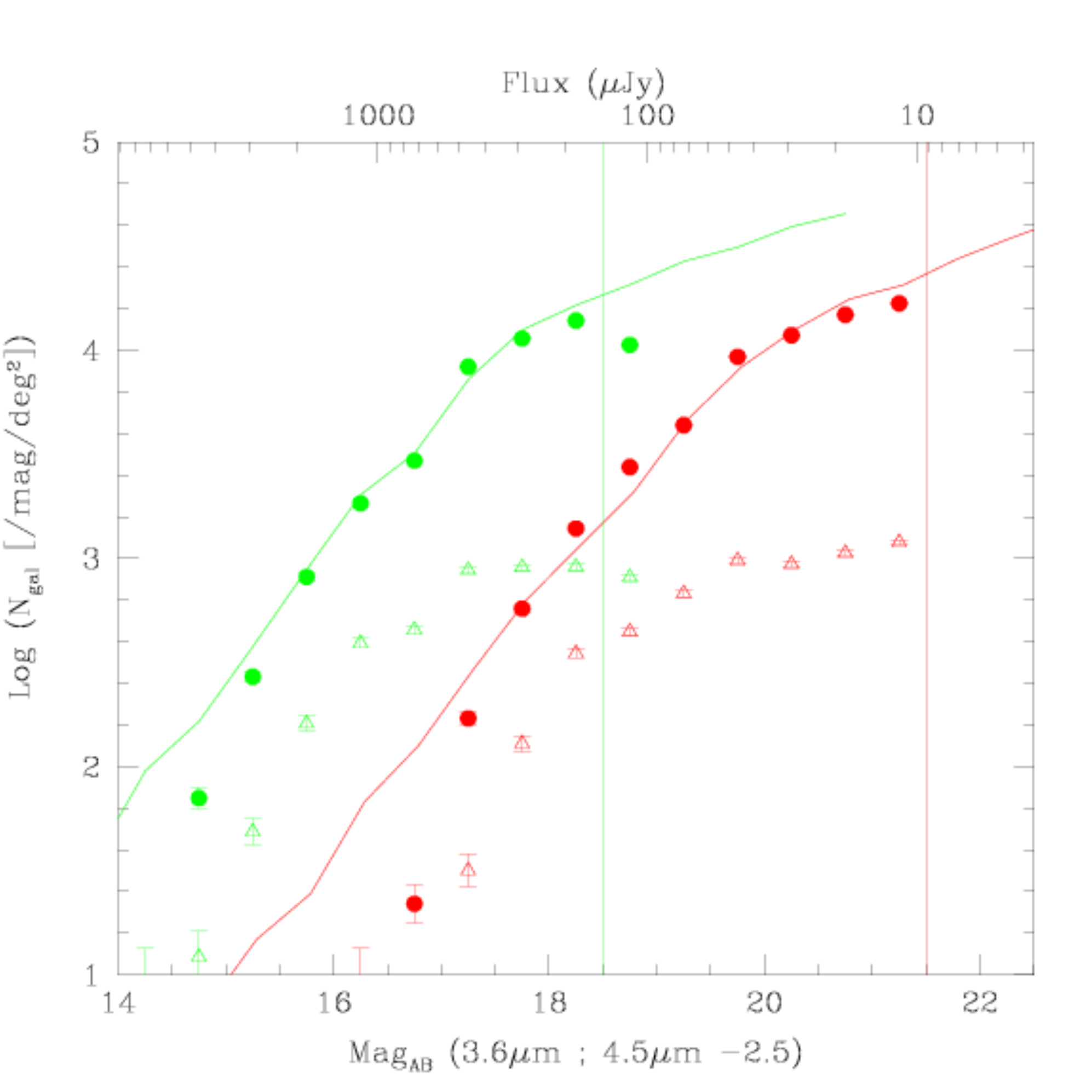}}
  \caption{Stars (triangles) and Galaxy (circles) number counts for the 3.6$\mu m$ (red symbols) and 4.5$\mu m$ (green symbols) samples. Solid lines show the IRAC number counts from Fazio et al. (2005).    }
  \label{fig:nm}
\end{figure}
%
  In this work we make use of the large amount of data collected in the 
 VVDS-0226-04 field.  The present sample is based on a 3.6$\mu m$ 
 flux limited sample from the SWIRE survey which overlaps the deep
 multi-colour imaging survey from the CFHTLS.
 The common area  between the two surveys corresponds to 
 0.85 square degree. We complete this dataset by including the deep
 spectro-photometric data  from the VVDS survey  and the infrared data 
 from the UKIDSS survey.  
\subsection{The SWIRE sample}
 The present sample corresponds to a $3.6\mu m$ flux limited sample with
  F(3.6$\mu m$)$\ge 9\mu Jy$ (or $m_{AB}(3.6)\le$21.5), based on the SWIRE 
  survey (Lonsdale et al., 2003). 
  The SWIRE  photometry  is based on the band-merged catalog
   including 3.6, 4.5, 5.6, 8.0$\mu m$ and 24$\mu m$ passband (Surace et al., 2005). 
   with a  typical $5\sigma$ depth  of 5.0, 9.0, 43, 40 and 311 $\mu$Jy respectively 
  We use the flux measurements derived in 3 arcsec apertures for faint sources 
   as suggested by Surace et al. (2005), while we adopt the adaptive 
   apertures (Kron magnitude;  Bertin and Arnouts, 1996)
  for the bright sources ($m_{AB}(3.6)\le 19.5$).   
  In Figure~\ref{fig:nm}, we show the IRAC 3.6 and 4.5$\mu m$ number counts.
  When comparing with the number counts from Fazio et al.  (2005),  the 
  3.6 $\mu m$ galaxy number counts are $\sim$80\% complete  at 9$\mu Jy$.
  At this depth within the common area with CFHTLS, we have $\sim$25500 sources.  
 \subsection{The  CFHTLS data} 
  The deep multi-colour photometry (u*g'r'i'z') from  
  the Canada-France-Hawaii Telescope Legacy Survey 
  is based on the T0003 release (CFHTLS-D1).
  These data cover one square degree, with sub-arcsecond seeing in all bands 
  and reach the limiting magnitudes (corresponding to 50\% completeness) of
  26.6, 26.5, 26.0, 26.0 and 25.2 in u*, g', r', i', z' respectively. A full description
  of the data will be presented in a forthcoming paper by Mc Cracken et al. (in prep.). \\  
  Thanks to the very deep optical data, almost all the IRAC
 sources have an optical counterpart, except a negligible fraction ($\sim 0.5\%$), 
  due to  unmatched positions within  1.5arcsec.
 \subsection{The VVDS data}   
  The VIMOS VLT Deep Survey consists of  deep photometry  and 
  spectroscopy (Le F\`evre et al., 2005):\\ 
 $\bullet$  Deep B,V,R,I imaging with a depth (50\% completeness) of
  26.5, 26.2, 25.9, 25.0,  respectively (Mc Cracken et al., 2003). 
  In addition, J and K observations with NTT-SOFI have been obtained at
  the depth (50\% completeness) of 24.2 and 23.8, respectively over
  172 arcmin$^2$ (Iovino et al., 2005). \\
 $\bullet$ The first epoch VVDS spectroscopic sample is based on a randomly
  selected sample of $\sim 9000$ sources in the magnitude range $17\le I \le 24.0$
  (Le F\`evre et al., 2004).  The spectroscopic area overlapping the
  SWIRE-CFHTLS  data is 0.42 deg$^2$ and provides $\sim 1500$ 
  secure spectra for sources with $m_{AB}(3.6\mu m)\le 21.5$.   
\subsection{The UKIDSS data}
 Finally, we complete our dataset with the J and K photometry from the 
 UKIDSS Ultra Deep Survey  (Lawrence et al, 2006) 
  based on the DR1 release (Warren et al., 2007). 
 These data reach a depth (5$\sigma$ limits) of 22.5 and 22.0 
 in J and K, respectively.  The overlap with SWIRE-CFHTLS is  0.55 and 0.76 deg$^2$ for J and  K  bands, respectively.  We match the UDS sources with the optical data
 within a 1 arcsec radius. Within the common areas,  89\% (or 93\%) of the 3.6 
 selected sources have a J (or K) flux measurement. 
\subsection{The combined sample}
\begin{figure}
 \resizebox{\hsize}{!}{\includegraphics{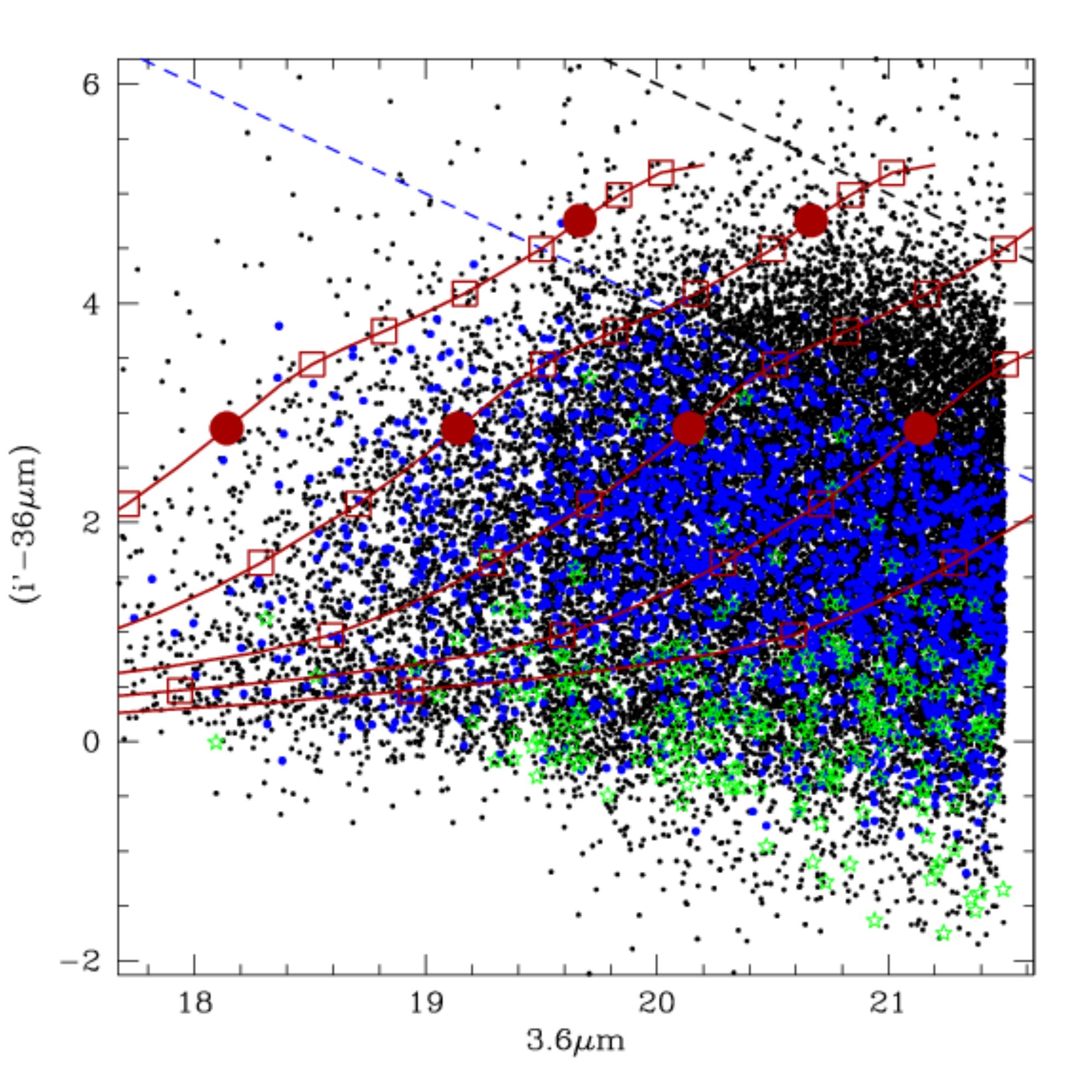}}
 \caption{Color-magnitude distribution with $(i-3.6\mu m)$ vs $3.6\mu m$.
 We show the  spectroscopic sample for galaxies (filled blue  symbols)
 and for stars (green stars symbols). The spectroscopic flux limit ($I\le 24$)
 and the $5\sigma$ detection limit for the CFHTLS ($i'\sim 26$) are shown as
  dashed lines  (lower and upper line respectively).  
 The solid lines show the behaviour of an elliptical galaxy (with $\tau=0.1Gyr$ 
 and $z_{form}=6$)  moving from $z=2.5$  to $z=0$ (from red to blue color)
 and for different absolute K magnitudes between $K_{ABS}=-25$ (left solid line) and 
 $K_{ABS}=-21$ (right solid line). The large circles denote the redshifts:
 $z=1.0,2.0$, while open squares are spaced by $\delta z=0.2$.        }
 \label{fig:cm}
\end{figure}

 In Figure~\ref{fig:cm},  we show the $(i-3.6\mu m)$ vs $3.6\mu m$ 
 color-magnitude diagram for the whole sample (small dots).
 The galaxies and stars from the spectroscopic sample are shown 
 as blue and green symbols respectively. 
 We plot the  cut-offs introduced in the  spectroscopic sample due to the
 magnitude limit, $I_{AB}\le 24$ (lower dashed line) and in the
 CFHTLS by adopting the $5\sigma$ detection limit, $i'\sim 26$
 (upper dashed line). The evolutionary tracks of an
 elliptical galaxy,  formed at  $z=6$ and  with an e-folding parameter $\tau=0.1Gyr$ 
 (using the PEGASE model; Fioc et al., 1997) are shown as solid lines 
 for different K band absolute magnitudes ranging from $K_{ABS}=-25$
 to $-21$. While an optically selected sample with  $I\le 24$, 
 can detect ellipticals brighter than  $K_{ABS}\ge-23.0$  up to $z\sim 1.2$,  a 
 $3.6\mu m$ selected sample with $m(3.6\mu m)\le 21.5$ can detect them 
 up to $z\sim 2$ if very  deep optical photometry is available.
 The present sample is therefore well adapted to investigate 
 the evolution of old and/or massive galaxies up to $z\sim 2$, 
 providing a major step with respect to previous analyses
 based on optical selection.\\

%
\section{Photometric redshifts} 
\subsection{The method}
 We measure the photometric redshifts and we classify the whole population 
 in  galaxy/quasar/star based on the $\chi^2$ fitting analysis of the spectral  energy 
 distributions, using the photometric redshift code
 "Le Phare"\footnote{http://serweb.oamp.fr/perso/arnouts/LE\_PHARE.html}.
 In this work, we adopt a similar procedure to the one described by Ilbert et al. (2006). 
 We use empirical templates based on the four observed spectral types from
 Coleman et al. (1980) and add two starburst templates from Kinney et al. (1996).  
 Templates are extrapolated into ultraviolet and infra-red wavelengths using the GISSEL
 synthetic models (Bruzual and Charlot, 2003) and we refine the set of SEDs with a
 linear interpolaton amongst  the original SEDs.
 For Scd and later spectral types, we allow for different amounts of dust attenuation with 
 a reddening excess E(B-V) varying from 0 to 0.6 and an interstellar extinction law  
 from Prevot et al. (2004). The spectroscopic sample is used to perform a template  
 optimization and estimate possible systematic shifts amongst the different passbands
 as introduced by Ilbert et al. (2006).
 The SED fitting is performed from the u* to the 3.6 and 4.5$\mu m$ photometric
 passbands (the two latter passbands being still dominated by stellar light). 
 All objects have at least 4 passbands with measured fluxes which ensures reliable
 SED fitting analysis. \\
%
\begin{figure}
  \includegraphics[width=8.5cm]{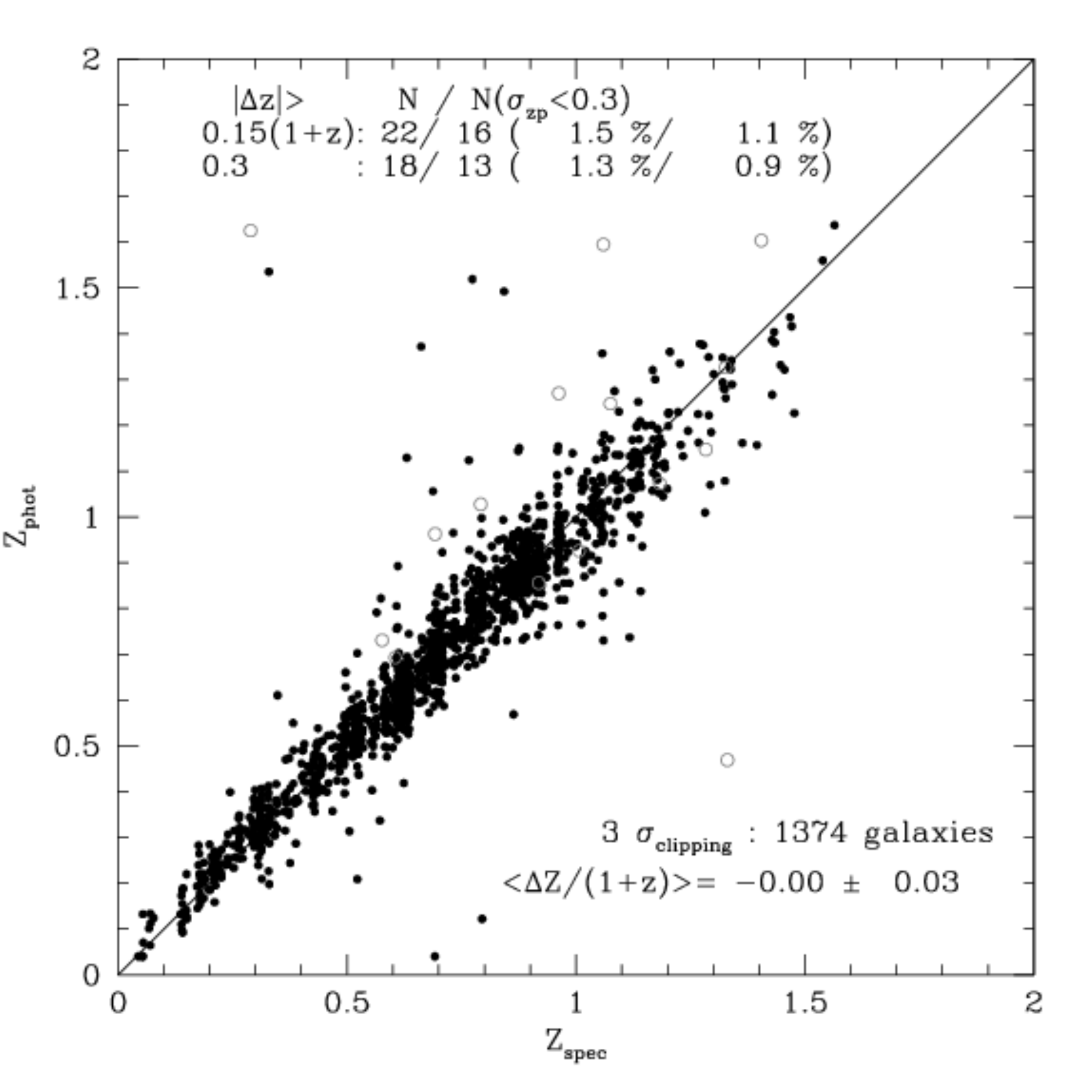}
  \caption{Photometric vs Spectroscopic redshift comparison,
   for $\sim 1400$ secure VVDS spectroscopic redshifts. Open symbols
   refer to objects with internal photo-z uncertainties  $\sigma(z_{phot})\ge 0.3$. } 
  \label{fig:zs}
\end{figure}
%
 In Fig~\ref{fig:zs}, we compare the photometric and spectroscopic
 redshifts for 1400 galaxies observed by the VVDS. 
 We obtain an  accuracy in the photo-z of  $\sigma[\Delta z/(1+z)]\sim 0.031$ 
 with no systematic shift and  a small fraction of catastrophic errors (1.5\% with
 $|\Delta z| \ge 0.15(1+z)$).   \\
 We consider as stars and QSOs (or AGN) the objects with a high SExtractor 
 stellarity index parameter in i' band (CLASS\_STAR$\ge 0.97$) and for which 
 either the star or the QSO template  provide a best $\chi^2$ fitting. 
%
 After removing these stars and QSOs, we end up with  a total sample 
 of $\sim$21200 galaxies out of 25500 sources.  \\
%
\begin{figure}  
  \includegraphics[width=8.5cm]{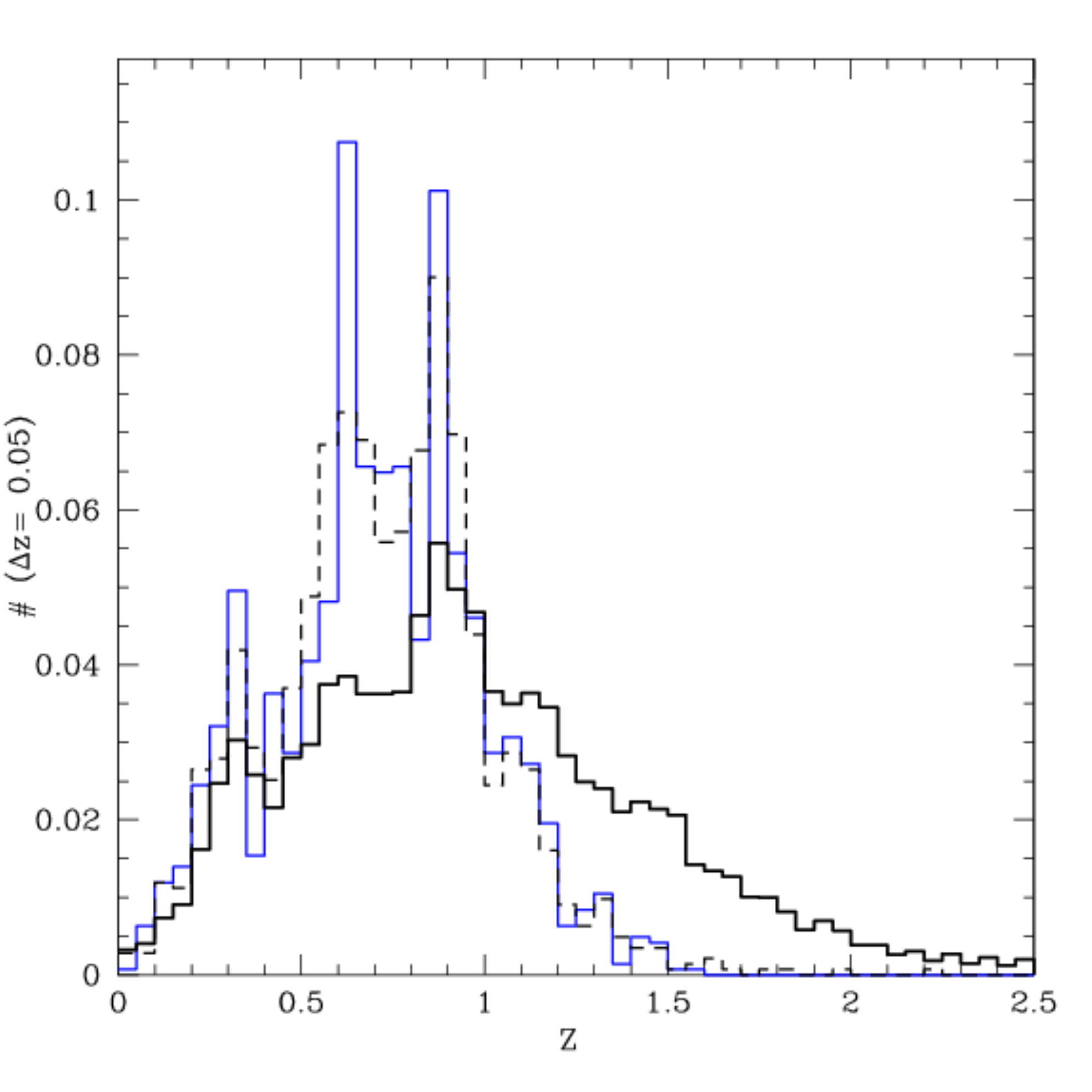}
  \caption{The redshift distributions for the $3.6\mu m$  sample 
   with $F(3.6\mu m)\ge 9 \mu Jy$, for the full photo-z sample  (solid thick line), the
  spectroscopic sub-sample (solid thin line) and 
  the photo-z of spectroscopic objects (dashed line).
  All three curves are normalized to unity.}
  \label{fig:nz}
\end{figure}
%
%
\subsection{The Redshift distributions}
\begin{figure}
 \resizebox{\hsize}{!}{\includegraphics{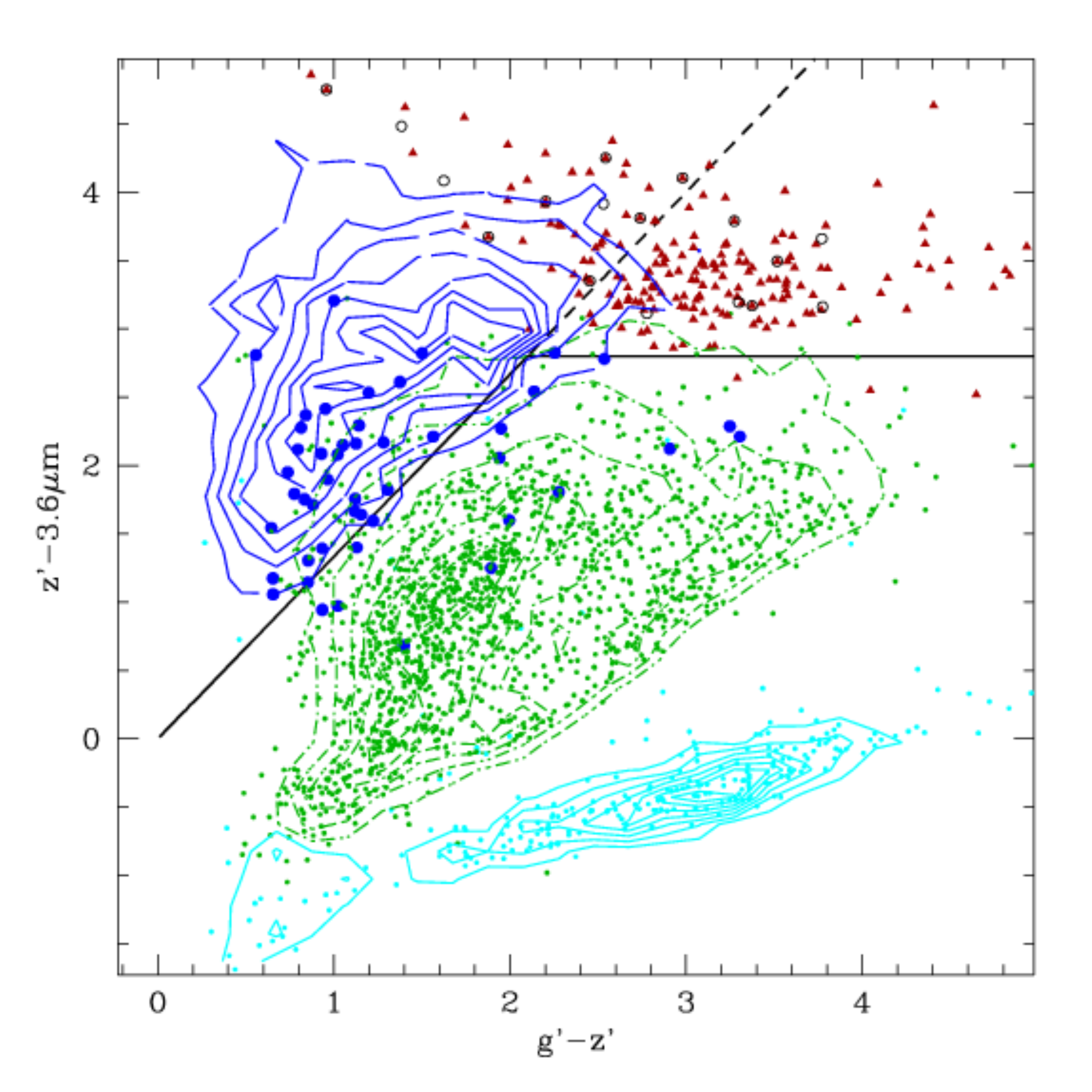}}
 \caption{Color-color diagram with  $(z'-3.6\mu m) vs (g'-z')$ for $3.6\mu$ sample.
 The filled dots show the spectroscopic objects while contours show the distributions of the photo-z sample for  $z\le 1.4$  (dashed-dotted  green contours), $z\ge 1.4$ for the  active/blue (long dashed blue contours) and  stars  (thin cyan contours).  
The  quiescent/red populations  with $z_{phot}\ge1.4$ are shown as triangles 
and the open circles show dusty galaxies observed at $24\mu m$  (F(24$\mu m) \ge 300\mu$Jy) and falling above the  $(NUV-r')_{ABS}$ criteria described in the text.
 The lines show a typical separation between the
 three classes in a similar way as  the BzK selection.
 }
 \label{fig:cc}
\end{figure}

 The redshift distributions are shown in Figure~\ref{fig:nz} for the whole sample 
 with  photometric redshifts  (thick line),  for the spectroscopic sub-sample
 ($I\le 24$, thin line) and for the photo-z's of the spectroscopic objects (dashed line). 
 The two latter distributions, both optically selected, are in excellent agreement, but 
 do not show the high redshift tail (up to z=2-2.5) observed for the
 3.6$\mu m$  sample.  We find that $\sim40\%$  of galaxies  with 
 $F(3.6\mu m)\ge 9\mu Jy$  are at $z\ge 1$.  A similar fraction, $\sim 35\%$,
 is obtained by Franceschini et al. (2006) for galaxies with
 $F(3.6\mu m)\ge 10\mu Jy$ in the GOODS-CDFS  field. Note however 
  that both samples rely  partially  on photometric redshifts.      
 Since the high-z tail ($z\ge 1.4$) cannot be verified
 by our spectroscopic sample,  we  have investigated the reliability of the 
 redshifts for  this high-z population through  color-color diagnostics. 
 In  Figure~\ref{fig:cc}, we show the (g'-z') vs (z'-36$\mu$m) color-color 
 diagram  as an indicator for galaxies at $z\ge 1.4$. This is similar to the 
 BzK diagnostic proposed by Daddi et al (2004). The locus of  galaxies 
 with $z_{phot}\ge 1.4$  is well separated from the low redshift population
 as illustrated by the VVDS spectroscopic sample.   This was also reported
 by Daddi et al. (2004) for the spectroscopic follow-up of BzK candidates.
 We conclude that our MIR selection at 3.6$\mu m$ allows us to probe
 a significant population of galaxies in the redshift range  $1\le z\le 2$ 
 with no indication of any bias from the lack of spectroscopic redshifts. \\

 The spectroscopic redshift distribution also shows strong peaks
 at $z\sim$ 0.35, 0.6 and 0.85. Two of them (z=0.35 and 0.85) are still
 observed  in the photo-z distribution although the area is two times 
 larger and the distribution is somewhat smeared by the photo-z errors. 
 These two peaks have been identified as large scale structures extending
 through the entire field  (Marinoni et al., 2005; Adami et al., 2007) and  we estimate 
 the  number density  to be in excess  by $\sim$20-25\% in these 
 structures (based on a smoothed redshift distribution derived by applying a 
 moving window with a width of $\Delta z=0.2$).  
 Despite the large field of view,  cosmic variance
 remains a significant  source of uncertainty as previously pointed out
 by Bell et al. (2004) in the case of the COMBO17 survey. 
 Following the recipes described by  Somerville et al. (2004), we
 estimate  the cosmic variance from large scale structure
 fluctuations in the $\Lambda$CDM paradigm.  In the low and high
 redshift bins (with volumes of 2.7 and 12.5 $10^5 Mpc^3$ respectively), 
 we expect  the cosmic variance to  produce density variations of 
 30\% and 10\% respectively, which appears to be  in good agreement with 
 our own estimates.  We decide to account for this effect as an additional 
 source of uncertainties  in the  measurements presented hereafter. 
%
%
\subsection{Estimates of absolute magnitudes}
 Throughout this paper we use absolute luminosity quantities. 
 In practice, to derive the absolute magnitudes we use an adaptive filter 
 method as introduced and tested on simulations by  Ilbert et al. (2005, Appendix A). 
 To reduce the uncertainties in the k-corrections used to derive the 
 absolute magnitude at $\lambda_{rest}$, we choose the apparent magnitude 
 in  the  filter ($\lambda_{obs}$) that best matches the redshifted  $\lambda_{rest}$.
 In the case where $\lambda_{obs}=\lambda_{rest}*(1+z)$, the k-correction is reduced to its redshift component ($2.5*log(1+z)$) with no template dependency.  \\
 All absolute magnitudes are derived according to this scheme except for the 
 specific case of the K absolute magnitude ($K_{ABS}$). We always use
 use the $3.6\mu m$ magnitudes for $z\ge 0.4$. This choice yields null
 k-correction at $z\sim 0.6$ and requires small k-correction in highest
 redshifts with little template dependency. When available, at $z\le 0.4$,  
 we use the observed K band. \\
 In the next section we use the $NUV$ and $r'$ absolute luminosities.
 Over the redshift range $0.3\le z \le 1.75$ the rest-frame $NUV$ ($\sim 2350\AA$)
 and $r'$ passbands correspond to observed wavelengths of 
 $0.3\mu m \le \lambda \le  0.65\mu m$ and 
 $0.8\mu m \le \lambda \le 1.7\mu m$ respectively. 
 We therefore need only small extrapolations for the $NUV$ absolute
 magnitudes in the lowest z-bin.  For the $r'$ filter, the near infrared photometry 
 is required at  high redshift, which is available for a large fraction of our sample.
 We note however, that  even without near-infrared photometry  the shape of the SED
 is always well constrained in the NIR domain, because all our objects have 
 per definition a measurement at $3.6\mu m$. We therefore obtain a good estimate 
 of the $r'$ absolute magnitudes even at high redshift.  \\ 
 We have derived  the absolute magnitudes  based on the empirical SEDs used to  
 measure the photo-z.  As a external check, we have also measured the
 absolute magnitudes  based on a library of SEDs from PEGASE2  code (Fioc et al., 1997).
 Thanks to the use of the adaptive filter method, the results from the two libraries
 are fully consistent  with  an r.m.s. dispersion for the  $(NUV-r')_{ABS}$ color of $\sigma[\Delta(NUV-r')]\sim 0.2$. This is  small compared to the large dynamical range of this color (see Fig~\ref{fig:nuv_r}).
\subsection{The selection of quiescent galaxies}
\label{sec:color}
%
\begin{figure}
\resizebox{9cm}{!}{\includegraphics{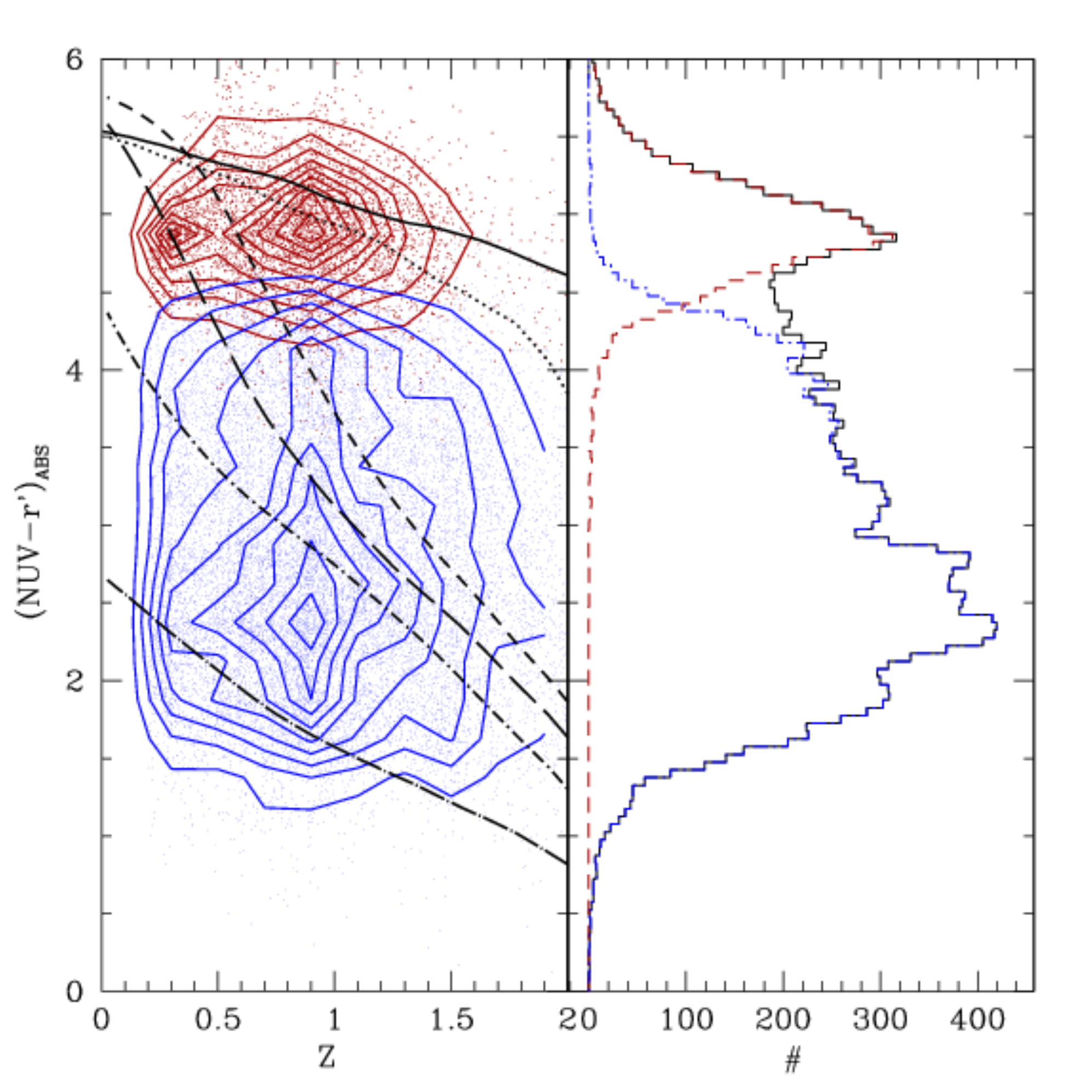}}
\caption{Left panel:  $(NUV-r')_{ABS}$ color vs redshift 
 for the quiescent/red (red contours) and active/blue (blue contours) 
 populations. Color evolution for models with z$_{f}=3, 6$  and $\tau=0.1$
  (dotted and solid lines resp.); 
  z$_{f}=3$  and   $\tau=1, 1.5$  (short and long dashed lines);  
 $\tau=2,\infty$ (dot-short dashed and  dot-long dashed lines resp).
 Right panel: histogram for the  $(NUV-r')_{ABS}$ color (solid line).
 The dashed lines show the distributions of the  
 two populations as separated by SED templates (see text).}
\label{fig:nuv_r}
\end{figure}
%
%
\begin{figure}
\resizebox{9cm}{!}{\includegraphics{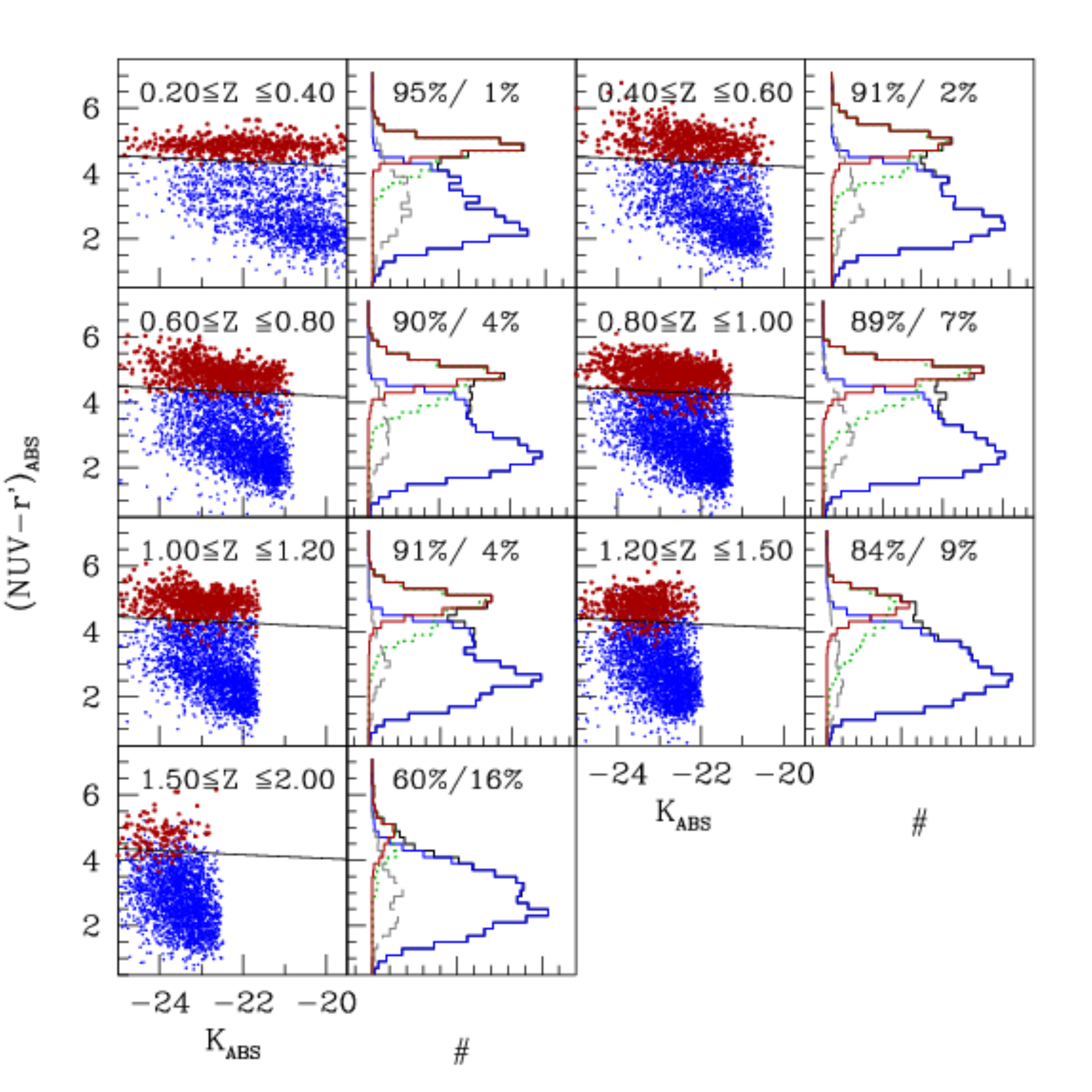}}
\caption{$(NUV-r')_{ABS}$ color vs $K_{ABS}$ distribution (left side) 
 as a function of redshift (as specified in each panel). 
 The two colored set of points correspond to the quiescent and 
 active galaxies as separated by SED templates. The solid lines denote the valley
 between the two populations as seen in the projected histograms (right side).
 The green dotted histograms show the selection of red galaxies based on  
  $U-B$ rest-frame color (see text).    
 The dashed histograms show the galaxies detected at 24$\mu m$ ($F(24\mu m)\ge 300\mu Jy$)}
\label{fig:nuv_r_k}
\end{figure}
%
 We split our sample in two main classes in order to distinguish 
 the red/quiescent  and blue/active galaxies.  To do so, we 
 explore and compare two approaches, where one is based on the color bimodality  
 and the other on an SED fitting procedure.   
 For the color bimodality we consider the excellent separation  
 provided by the $(NUV-r')_{ABS}$ color  (Figure~\ref{fig:nuv_r}; right panel) as
 suggested by recent GALEX results (Salim et al., 2005; Wyder et al., 2007).
 This is in contrast to the often used rest-frame $(U-V)_{ABS}$ color
 suggested by Bell et al. (2004).\\
 One of the advantages of the $(NUV-r')_{ABS}$ color is that the $NUV$ passband
 is sensitive to a stellar population with a light-weighted age of $<t>\sim10^8$yr
 while the $r'$ passband  is probing $<t>\ge 10^9$yr (Martin et al., 2005),  
 making the $(NUV-r')_{ABS}$ color  an excellent  indicator of the 
 current over past star  
 formation activity. Indeed, the GALEX-SDSS sample shows that $(NUV-r')_{ABS}$
 is tightly correlated with the birthrate b parameter ($b =SFR(t<10^8yr)/<SFR>$)
 with  $(NUV-r')_{ABS}\ge 4.3$  being associated to galaxies with  $b\le 0.1$ and
 morphologically with de Vaucouleur profiles (Salim et al., 2005).\\

 To illustrate  the redshift evolution in $(NUV-r')_{ABS}$, 
 we use the stellar synthesis model PEGASE2, with smooth star formation
 histories, described by different  e-folding times and formation redshifts
 ($\tau(Gyr)$, $z_{f}$).
 In Fig~\ref{fig:nuv_r},  left panel, we show the evolution of $(NUV-r')_{ABS}$ vs z
 for different SF histories ($\tau$)  and redshift of formation.
 Galaxies with $\tau \le 0.1$ Gyr are in the red clump at all z, 
 while models with higher $\tau$ move  progressively into the red sequence 
 at decreasing  redshifts (z$\sim$0.7 and 0.4 for $\tau=1$ and 1.5 Gyr respectively).
 All models with $\tau\ge 2 Gyr$ have  ongoing star formation 
 that prevents them from reaching the red sequence unless  their star formation 
 is quenched prematurely. 
 Therefore the  $(NUV-r')_{ABS}$ color bimodality  (Fig~\ref{fig:nuv_r}, right 
 panel) appears to be a  good way to separate active and quiescent galaxies 
 according to their birthrate parameter, with $b\le 0.1$ for quiescent galaxies.
 
\begin{figure}
  \resizebox{\hsize}{!}{\includegraphics{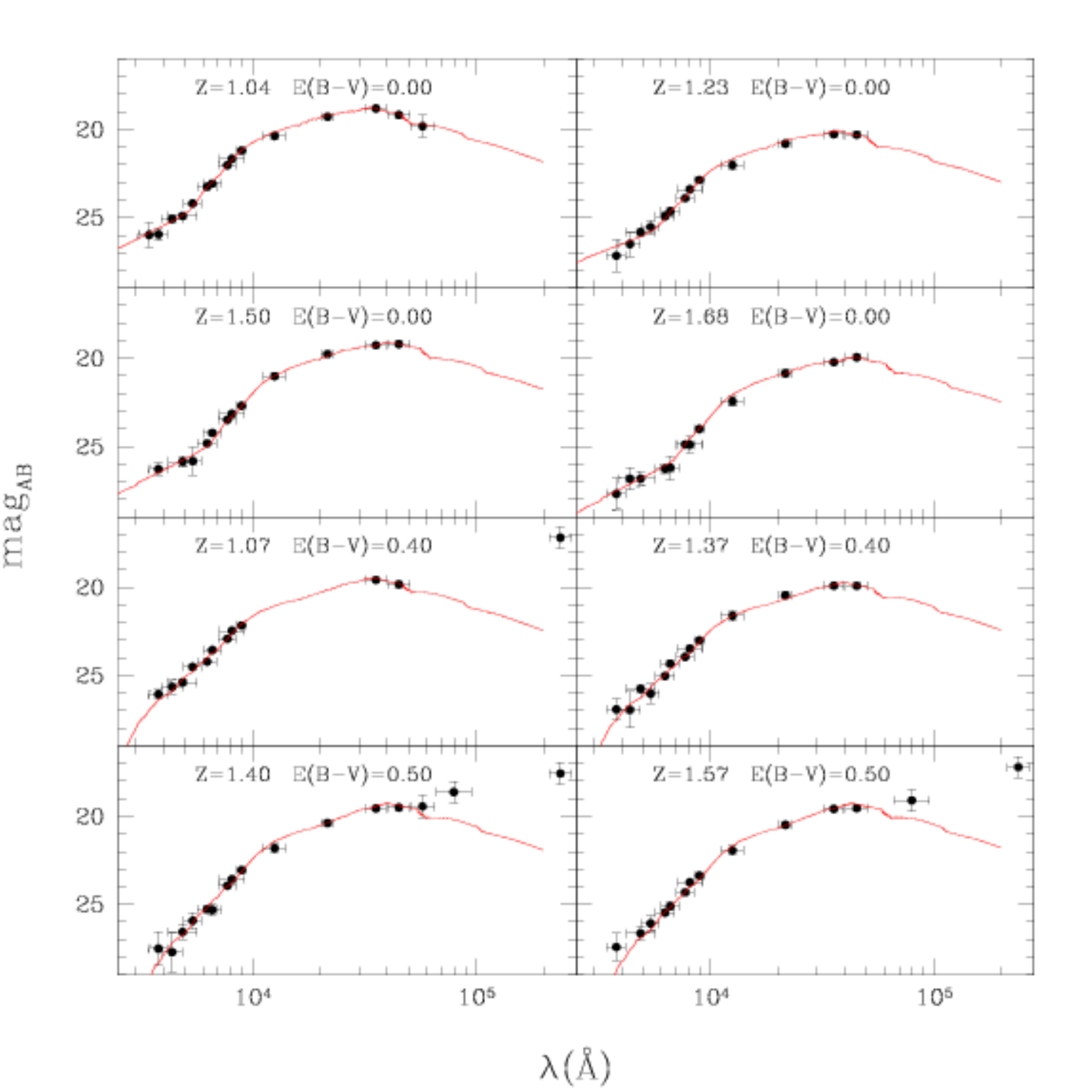}}
  \caption{ Examples of  SEDs for a sample of high-z galaxies 
 with  $(NUV-r')_{ABS}$ above of the color criterion (see text).  The fit is performed 
 only in the wavelength domain  $0.3\mu m \le \lambda \le 4.5 \mu m$.
 The top four objects are best fitted by an old/passive elliptical, while the lower
  four galaxies are best fitted by dusty star-forming SEDs.}
  \label{fig:sed}
\end{figure}

 In Figure~\ref{fig:nuv_r_k}, we show the galaxy distribution in the 
 diagram  $(NUV-r')_{ABS}$ vs $K_{ABS}$  and the projected histograms 
 for the different redshift bins. A strong bimodality  is observed at almost  all redshifts 
 providing a natural way to split the sample. 
  We define an adaptive color cut-off located in the valley and characterized by:
  $(NUV-r')_{ABS} = -0.06 \times (K_{ABS}+22) + b(z)$, where 
  $b(z)=4.4 - 0.13 \times z$ (solid lines). 
 However, some uncertainties remain on the nature of the population in the  
 red sequence  due  to the contamination by dusty  star-forming galaxies.
 A reddening excess of  $E(B-V)=0.2$ produces a color reddening of 
 $\Delta(NUV-r)_{ABS}=0.94$ (or 1.04), assuming starburst (or SMC) 
 extinction curve. We also show in Fig~\ref{fig:nuv_r_k}, the locations of those
 galaxies that are most likely dusty star-forming galaxies,
 as indicated by their detection at 
  $24\mu m$  (with $F(24\mu m)\ge 300\mu Jy$). 
  Between 5 to 9\% (depending on $z$) of this bright 24$\mu m$ population
  lies above our adopted colour cut and is a source of contamination for 
   our quiescent population. \\
 We then take a different approach,  by classifying  the galaxies
  according to the best-fit SED, on the basis of the  six original templates
  (but allowing for dust extinction for the later types), in a similar way as
  Zucca et al. (2006). We consider as quiescent galaxies only 
 those classified as elliptical  and all other galaxies as active.\\
 In Figure~\ref{fig:nuv_r_k}, we compare the SED-based classification with the  
 $(NUV-r')_{ABS}$ color distribution (quiescent as red points; active as blue points). 
 To quantify the overlap between the two different classifications, 
 in each panel, we measure the fraction of ellipticals in the sample
 above and below the color cut  previously defined.  We find that  the 
 SED-ellipticals dominate (with $\sim$ 90\%) the red peak population up to 
 $z\le 1.5$ with only a small  fraction falling below the cut.
%
 At $z\ge 1.5$,  the fraction of ellipticals above the cut represents only    
 60\% of the reddest objects while the remaining 40\% are best classified 
 as dusty star-forming galaxies.  In Fig~\ref{fig:sed}, we show examples of 
 the two types of SEDs  (Ell or dusty star-forming)  for galaxies above 
 the $(NUV-r')$ cut and  with $z_{phot}\ge 1.0$.  \\
%
%
  In Fig~\ref{fig:cc}, we show the distribution of  SED-elliptical galaxies with $z\ge 1.4$
  (red triangles).  As for the BzK selection with a spectroscopic validation
  (Daddi et al., 2004), they are reasonably well separated  from the active galaxies 
   with $z\ge 1.4$ (blue contours).   We also show the bright $24\mu m$ population
  with $(NUV-r')_{ABS}$ in the red sequence and $z\ge 1.4$ (open circles).
  They are spread  all over the upper part of the diagram. Amongst this population, 
  $\sim$50\% of them has been correctly adjusted by dusty star-forming SEDs. 
  This shows that SED fitting can get rid partially of the dusty galaxies falling in the red 
  sequence.  As  a last test to quantify the residual contamination of the red/quiescent 
   sample,  we project our elliptical sample in the $(J-K)$ vs $(r'-K)$
  color diagram. This is susceptible to distinguish the early-types from the dusty
  starbursts (Mannuzzi \& Pozzetti, 2000;   Daddi et al., 2004).
  We find that $\sim$15\% of the sample lies in the dusty starburst locus. If this color  
  separation were to be efficient,  this would  mean  that our quiescent sample may
  still be moderately contaminated by dusty galaxies.
 
 In conclusion, we have analyzed  two methods to select the
 red/quiescent galaxies based either on the $(NUV-r')_{ABS}$ color or   
 on the selection of quiescent galaxies according to SED fitting. 
 We have shown that the SED-quiescent population represents the large
 majority of galaxies in the red peak of the bimodality distribution.  
 We decide to adopt the SED-fitting classification because it allows to reduce
  the contamination of dusty starbursts in our red/quiescent  sample.
  At $z\ge 1.4$, we estimate the residual contamination to be around
  $\sim$15\%. Only deeper Spitzer-MIPS observations could allow to better 
  disentangle the two populations. However we note that the results
 discussed in the following sections are marginally affected by the
 chosen selection.\\
 %
 We end up with a sample of  4425 quiescent and 16770  active galaxies. 

%
%
\section{The stellar mass to light ratio (M/L$_K$)} 
\label{sec:ml}  
 We now derive the evolution of the stellar mass to 
 light ratio (M/L$_K$) with redshift for the active and quiescent populations.  
 This will allow us to assess stellar mass quantities. 
 Thanks to the large VVDS spectroscopic
 sample, we can measure how the M/L$_K$ varies with redshift. In particular, 
 the use of spectroscopic redshifts allows us to avoid mass 
 uncertainties relative to photo-z and we thus obtain accurate 
 estimates of the scatter for these relations.

 We derive the stellar mass $M^{\star}$  for each galaxy by fitting 
 the multi-band photometry.
 In the present work, we base ourselves on the 
 Bruzual \& Charlot (2003) stellar population 
 model. We have used this model to construct a library of star formation histories 
 that includes stochastic bursts (see Kauffmann et al. 2003 and Salim et al. 2005 for 
 similar libraries). Although this library uses a Chabrier (2003) IMF, 
 we have decided to adopt a classical Salpeter IMF throughout this paper.
 We therefore apply  a systematic shift of $+0.24$ dex to the masses 
 derived with a Chabrier IMF. In brief, our method uses the Bayesian 
 probability distribution functions as described in Appendix A of Kauffmann et al. 
 (2003) to derive parameter estimates for each observed SED, including errors 
 on each derived parameters, in this case the stellar mass.
 A full description of our mass  estimation routines 
 will appear in forthcoming papers (Lamareille et al. 2007, Walcher et al. 2007).
 For now we refer to previous works  
 for discussion on the accuracy of the stellar mass estimates
 by SED fitting methods, in particular the choice of star formation histories
 with or without episodic bursts  (Borch et al., 2006; 
 Pozzetti et al, 2007) and the addition of Mid-IR bands (Fontana et al., 2006). 
 We note, however, that our stellar mass estimates may be affected by the established 
 problems pertaining to the contribution of short-lived, luminous, infrared-bright 
 stars in intermediate age population ($\sim 1$ Gyr) (Maraston et al. 2006).
 Our mass estimates would possibly need to be changed by 0.1 to 0.2 dex 
 (see Pozzetti et al. 2007 for an in-depth discussion). A final solution 
 to this problem affecting all publicly available stellar population synthesis models 
 is however not imminent.

\begin{figure}
\resizebox{\hsize}{!}{\includegraphics{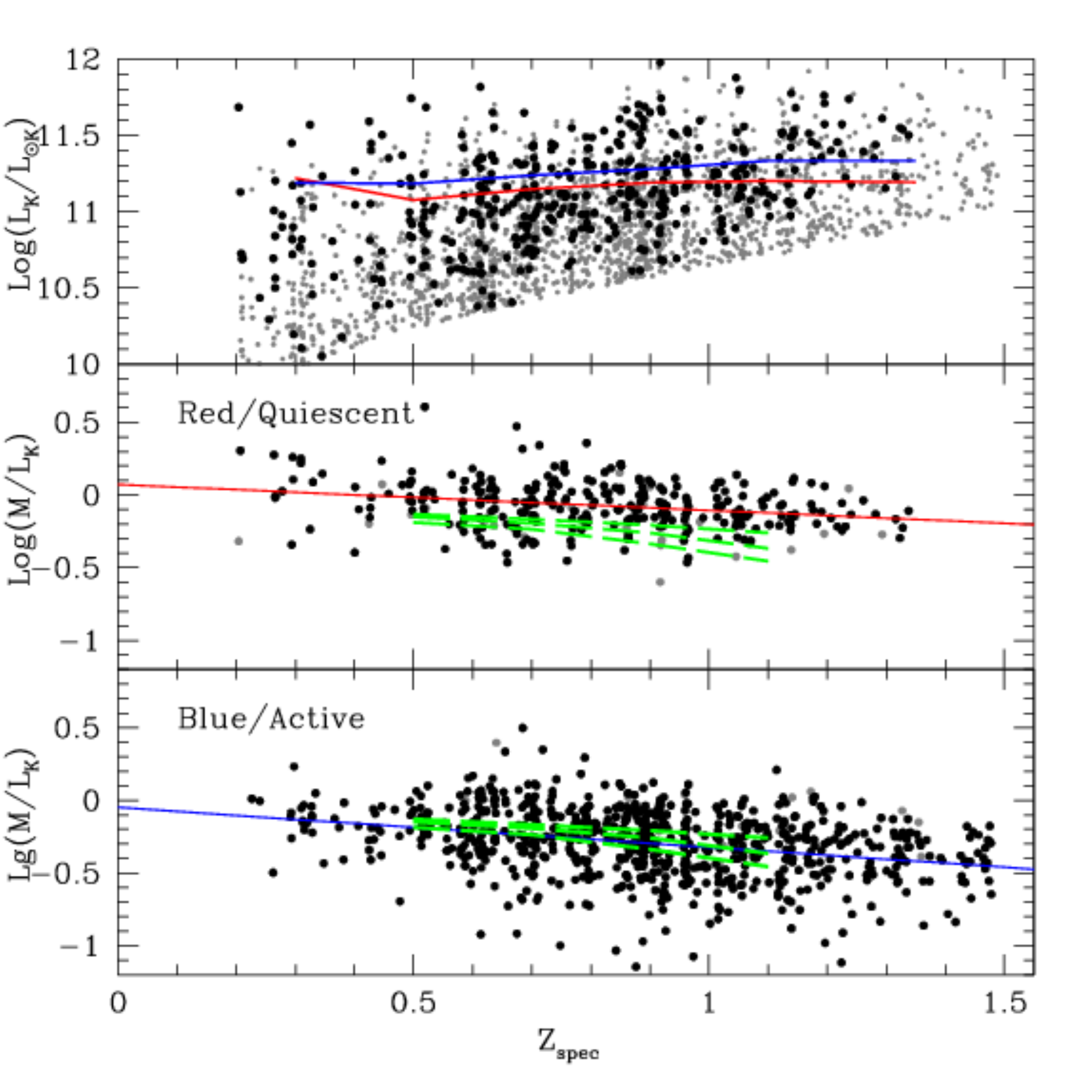}}
 \caption{Upper panel:  distribution of the K luminosity ($L_K/L_{\odot K}$)
 for the spectroscopic sample  (quiescent galaxies: black filled circles; 
 active galaxies: grey small circles).
 The characteristic luminosity $L^{\star}$  for the Red/quiescent and Active/blue
  samples are shown as solid lines. 
 Lower panels: Behavior of the Mass to light ratio (M/L$_K$) for the
 quiescent (middle panel) and active (lower panel) samples with the 
 best fit  shown as solid lines.  Comparison with results from 
 Drory et al. (2004; long dashed lines)  based on three different 
 stellar mass cuts: 
 $M/M_{\odot}\ge 2. 10^{11}$, 
 $10^{11}\le M/M_{\odot}\le2. 10^{11}$, and  
 $4.10^{10}\le M/M_{\odot}\le 10^{11}$. } 
 \label{fig:ml}
\end{figure}

 In Figure~\ref{fig:ml} (upper panel), we show the redshift distribution
 of the K luminosity  ($L_K/L_{\odot K}$) for the quiescent (large black circles)
 and active (small grey circles)  populations.  The solid lines show 
 the evolution of the characteristic luminosity, $L^{\star}$, derived below in 
 Section \ref{sec:lf} and reported in Table \ref{tab:lf}. The  distribution of 
 M/L$_K$ as a function of redshift for the quiescent and active 
 samples are shown in the middle and lower panels respectively. 
 Before deriving the M/L$_K $ versus redshift relations from the spectroscopic sample
 we have tested several subsets:
 The first one includes only objects brighter than $L_K \ge 10^{11}L_{\odot K}$
 to define an unbiased luminosity sample up to $z=1.5$.  The second one 
 considers objects with a luminosity between
 $0.4 L^{\star}(z)\le L_K(z) \le 2.5 L^{\star}(z)$, 
 allowing to probe the evolution of the M/L$_K$ for galaxies 
 around $L^{\star}(z)$ which  are the main contributors to the total luminosity
 density with 70\%, 45\% and 55\% for the quiescent, active and 
 total samples, respectively.  
 Because the spectroscopic sample is optically selected ($I\le 24$), 
 we have also considered a sub-sample with $m3.6 \le 20.5$ (which means
 one magnitude  brighter than the full sample)  to insure a  fair sampling of the 
 $(i'-3.6\mu m)$ colors at all redshifts. 
 All these subsamples are found to provide consistent results within the parameter's 
 uncertainties, therefore we only report the values  for the three samples
 with the selection centered  around their respective $L^{\star}(z)$.
  For the three samples, we fit the redshift evolution of the M/L ratio by a
 power-law defined as $\log<M/L_K>=a\times Z+b$,  with a global
 rms  given by   $\sigma$. \\
 For the global sample, based on 999 galaxies: 
 \begin{equation}
 a= -0.30\pm 0.03 , b=0.03\pm 0.03, \sigma=0.22
\end{equation}
For the active/blue sample, based on 753 galaxies: 
\begin{equation}
 a= -0.27\pm 0.03, b=- 0.05\pm 0.03, \sigma=0.21
\end{equation}
For the quiescent/red sample , based on 298 galaxies: 
\begin{equation}
 a= -0.18\pm 0.04 , b=+ 0.07\pm 0.04, \sigma=0.15
\end{equation}
 The fits to the quiescent and active samples are shown as solid
 lines in Fig~\ref{fig:ml}.
 All subsamples show a similar trend with a gradual decline of the $<\mbox{M/L}_K>$ 
 by a  factor $\sim$ 1.5 to 2 up to $z=1$. 
 Similar results were obtained by Drory et al. (2004) 
 for samples selected with different mass limits, based on the NIR
 MUNICS survey, and are shown in  Fig~\ref{fig:ml} (long dashed lines)
 The behavior  of the M/L is mainly driven by  the stellar activity.
 The youngest and bluest stellar population have the smallest M/L$_{\lambda}$
 (Bell and de Jong, 2001; Drory et al., 2004).  Similarly, the decline 
 of the M/L$_{\lambda}$ with redshift reflects the well established 
 decrease of the specific star formation rate (downsizing effect; Cowie et al.,
 1996:  for a galaxy  with same mass, the SF activity is higher in the past). 
 The smaller decline of M/L$_K$ with redshift observed for the quiescent sample 
 reflects  an older mean age of the stellar population, pushing the epoch
 of formation of the stellar component to high redshift.
 We find that the quiescent and active samples show a rather small scatter along this
 relation,  with only 0.15dex and 0.22dex scatter respectively. 
 The main origin for the scatter is the dependance of the M/L$_K$ 
 on the star formation history (SFH). The more complex SFH for the active population 
  could be responsible for the larger scatter. 
%
%
%
%

%
\section{The K band luminosity function and density} 
\subsection{The K rest-frame luminosity function}
\label{sec:lf} 
%
\begin{figure}
 \resizebox{\hsize}{!}{\includegraphics{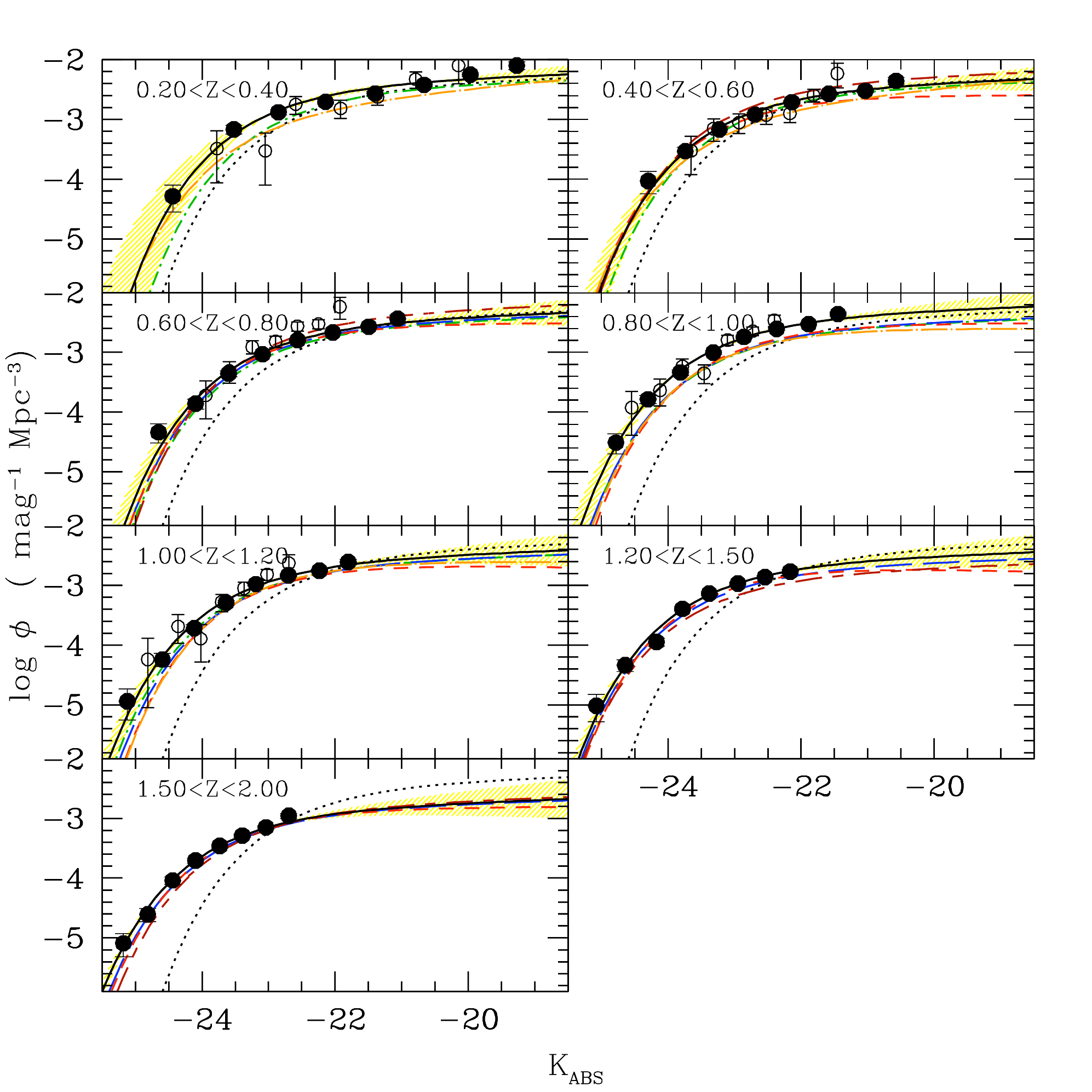}}
 \caption{
 K-band rest-frame Luminosity functions, in different redshift slices, for the 
 whole photo-z sample ($V_{max}$: filled dark circles, STY: solid lines and shaded area 
 based on slope uncertainties) and  the spectroscopic sample ($V_{max}$: open circles).
 Also shown are the local LF  from Kochanek et al. (2001; dotted line) and high-z LFs from Drory et al. (2003; green dahed-dot line), Caputi et al. (2005; blue long dashed line), Pozzetti et al. (2004; orange long dashed-dot line), Saracco et al. (2006; dark red long-short dahed line, $\alpha=-1.1$), Cirasuolo et al. (2006; red dashed line). }
\label{fig:lf}
\end{figure}
%
\subsubsection{K-LF measurements}
\begin{figure}
 \resizebox{\hsize}{!}{\includegraphics{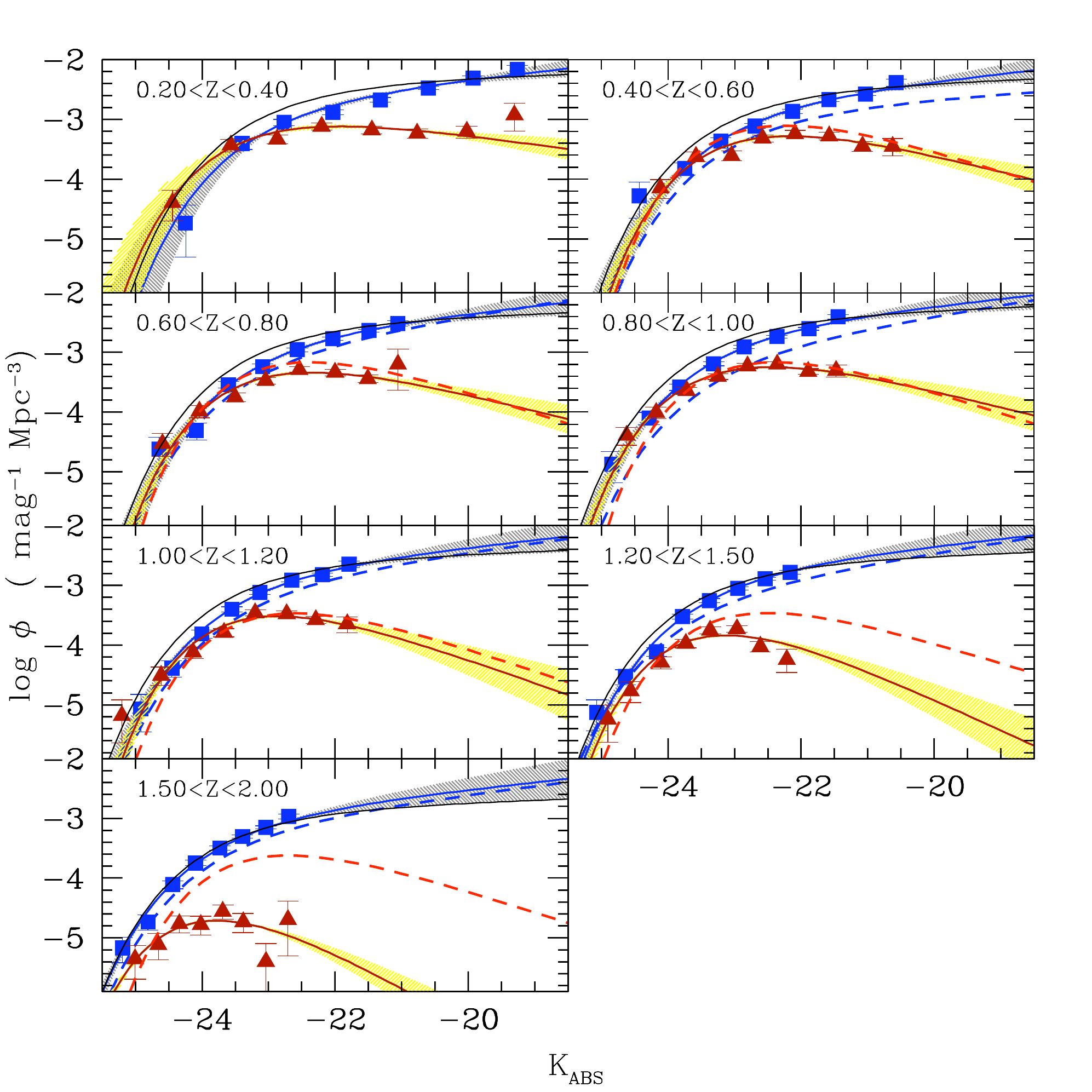}}
 \caption{ K-rest luminosity functions for the blue/active (blue squares and solid lines)
  and the red/quiescent (red triangles and solid lines) samples. The thin black lines
   refer to the global sample from Fig~\ref{fig:lf}.  We compare our LFs with  the red 
   and blue samples from the UKIDSS survey (Cirasuolo et al., 2006;  red and blue dashed lines).}
\label{fig:lft}
\end{figure}

 We measure the K rest-frame luminosity function (LF)
 by adopting the  $V_{max}$ and  STY  estimators  from the VVDS
 LF tool (ALF; Ilbert et al. , 2004).  In Figure~\ref{fig:lf}, we show the LFs  
 for the whole and spectroscopic samples and in Figure~\ref{fig:lft}  
 for the red/quiescent and blue/active samples. To measure the LFs, 
 a weight has been applied as a function of apparent magnitude to account
 for the incompleteness in number counts. 
 In the case of  the spectroscopic sample,  we derive the LFs 
 only for galaxies with  $m_{AB}(3.6\mu)\le 20.5$ 
 and $0.2\le z\le 1.2$.  This restriction is due to the I band selection effect
 (see Fig~\ref{fig:cm}).  An additional weight is applied to the spectroscopic
 galaxies to take into account the sampling rate of the VVDS
 observations  and its optical selection (as described in Ilbert et al., 2005).\\
%
%
In Table~\ref{tab:lf} we report the values of the STY parameters 
 for the different samples. The errorbars on the  parameters refer to the Poisson
  errors and slope uncertainties. 
 For the global and blue samples, we have fixed the faint-end slopes of the STY 
 estimator at low and high redshifts to the values observed between 
 $0.6\le z\le 1.0$, where we have the best constraints in the bright and faint-end
 simultaneously. This choice is consistent with local 
 (Kochanek et al. , 2001) and  high redshift analysis (Drory et al., 2004; 
 Caputi et al., 2005; Cirasuolo et al., 2006). 
 For the  red sample, we adopt a variable slope, 
 although  the low and high-z slopes are less well constrained. 
  At low-z ($z\le 0.4$) , we use  $\alpha=-0.6$, consistent with the slope
  observed  by Bell et al. (2004) from a local elliptical sample. 
 At $z\ge 1$, we adopt a gradual flattening of the slope that best reflects the
 evolution of  the $V_{max}$  estimator. However the depth of the current data 
 does not allow any statement on the reality of this apparent flattening and deeper 
 observations are required  to confirm it.  We note that the choice of the slope
 does not affect the discussion in next sections because  changes  to the estimates 
 of the luminosity densities are marginal 
 ($\rho_L(\alpha) - \rho_L(\alpha=-0.3)=-0.01,-0.02,-0.04$dex for $<z>=$1.1, 1.35, 1.75, resp.).
\subsubsection{Additional source of uncertainties }
 Additional sources of uncertainties on the LFs come from the photo-z and 
 cosmic variance. We have estimated the uncertainties relative to the photo-zs 
 from 100 mock samples based on the redshift probability distribution (PdZ) of each
 source (as in Bolzonella et al., 2000) and by recomputing the  type separation
 and luminosity functions. Considering only  the variations of $\Phi^{\star}$,  
 we find uncertainties in the density to be between 5\% to 12\% for the total
 and active samples, depending on the redshift bin. For the quiescent sample, the
 uncertainties vary between 7\% and 28\%.  This method allows to 
 account simultaneously for the large photo-z uncertainties of the ellipticals at high-z
 and to estimate the stability of  the quiescent vs active separation at high redshift. 
 We report  these uncertainties in Table~\ref{tab:lf} (($d\Phi/\Phi$)$_{PdZ}$).
 Except for the quiescent sample at high z, the  photo-z uncertainties
 are not a dominant source of errors in the present work.  
 
 It is more difficult to evaluate the effect of cosmic variance. 
 As a first attempt we have measured the global LFs in 10 non overlapping
 fields covering the whole area and find variations of $\Phi^{\star}$
 of between 0.07dex to 0.2dex.  
 However, the fact that the 10 fields are  located in a single area of the sky 
 prevents us from using them as cosmic variance uncertainties over the whole survey 
 (adopting a simple rescaling as $\sqrt{10}$). 
 Therefore, as mentioned in Section \ref{sec:data}, 
 we adopt the formal approach  discussed by Somerville et al. (2004). 
 The cosmic variance is estimated as  $\sigma=b \sigma_{DM}$, where  $\sigma_{DM}$
  is the variance of the dark matter that depends on the comoving volume 
  (their Figure 3, right panel)  and b is the bias relative to each specific sample
  that depends on the comoving number density (their Fig 3, left panel). 
 Theses uncertainties are reported in Table~\ref{tab:lf} (($d\Phi/\Phi$)$_{CV}$).
  Cosmic variance remains the dominant source of uncertainty for the whole and
  active samples at all redshift, while for the quiescent population 
  the photo-z uncertainty is the major source of uncertainty at $z\ge 1.35$. 
%
%
\subsubsection{Evolution of the K-band LFs and comparison with other studies }
In Fig~\ref{fig:lf}, we compare the global LFs with other studies. 
Beyond the fluctuations of $\Phi^{\star}$ due to the presence of large 
 scale structures, our LFs  appear slightly brighter than previous studies,
 by about 10\%. This is however consistent with the expected calibration
 uncertainty of the  SWIRE-IRAC flux (Surace et al., 2005).  
 Comparing $K^{\star}$  to the local measurement by Kochanek et al. (2001),
 we observe a brightening of  $\Delta K^{\star}\sim -0.8 (-1.0) \pm 0.2$ 
 from  $z=0$ to $z=1.5 (2)$.
 Regarding the comoving density parameter ($\Phi^{\star}$)
 we observe a decline by  a factor 1.7$\pm0.2$  up to $z=1.2$ and 2.5$\pm0.5$  up to $z=2.0$ as compared to local estimate.
 This mild luminosity brightening and modest density decline agree with
 previous studies (Caputi et al., 2005; Saracco et al., 2006, Cirasuolo et al., 2006). \\
  In Fig~\ref{fig:lft}, we compare the LF per types with our global LF and
 results from  UKIDSS based on  $(U-B)_{rest}$ color selection (Cirasuolo et al., 2006).  
  Our active sample shows a similar brightening  than the global sample. 
  The density, $\Phi^{\star}$, appears stable between $0.2\le z \le 1.5$
  and starts to decline at higher z.  
   For the quiescent sample a different behaviour  is observed. 
  Up to $z=1.2$, the density declines  by a factor 2$^{+1}_{-0.7}$,  
  followed by a sharp drop, by  a factor $16^{+9}_{-6}$, between  $1.2\le z \le 2$.  
  While the behaviour of the active and quiescent samples agrees with UKIDSS 
  results on blue and red galaxies
  (concerning faint end slopes and global evolution), we obtain  a different ratio 
  between the blue/active and red/quiescent populations  and they do not observe
  the strong decline for the red population at $z\ge 1.2$. We suggest two differences 
  in the selection criterion for red/quiescent galaxies as possible source to this  
  discrepancy:
 \begin{itemize}
 \item 
   In Fig~\ref{fig:nuv_r_k}, we show the  $(NUV-r')$  color distribution (green dotted  
   histograms)  for galaxies selected on the basis  of  the optical color $(U-B)$, in a
   similar way  as  Cirasuolo et al..  
  While the Cirasuolo selection includes our quiescent sample, it also includes 
  galaxies with  $(NUV-r)$ as blue as $2.5-3$. 
   This means that the optical criterion yields a number of red galaxies that is larger by 
  $\sim$40 to 50\% depending on redshift.  
   \item the modest decline of the red LF at $z\ge 1.2$ for UKIDSS  can also takes
    its origin in  the color criterion and in the inclusion of the larger fraction of red dusty 
   starbursts at high-z.  
    Regarding our selection of a quiescent sample, we were conservative in the sense 
  that we required that the galaxies do not reveal signs of 
  star formation activities according to their $(NUV-r')$ color and by using  
  multi-color SED fitting to exclude dusty starbursts.  
   In our red population at  $1.5\le z \le 2$, only  60\% of 
   galaxies is considered as quiescent, the rest being dusty starbursts.
   We expect this effect to be even stronger in the case of an optical  
   color selection. 
  \end{itemize}
 The strong evolution between $z=2$ and $z=1.2$ of the LF parameters of the 
 quiescent sample alone
 suggests that we are probing the epoch when an increasing number of 
 galaxies stop their star formation activities and  turn quiescent.
 However, the LF parameters are strongly correlated with each others and their 
 interpretation shaky. We therefore now turn to discussing the evolution of the
 different samples via the measurements of integrated 
 quantities (luminosity densities and stellar mass densities).
%
%
%
%
\subsection{The K band luminosity density} 
\label{sec:ld} 
\begin{figure}
 \resizebox{\hsize}{!}{\includegraphics{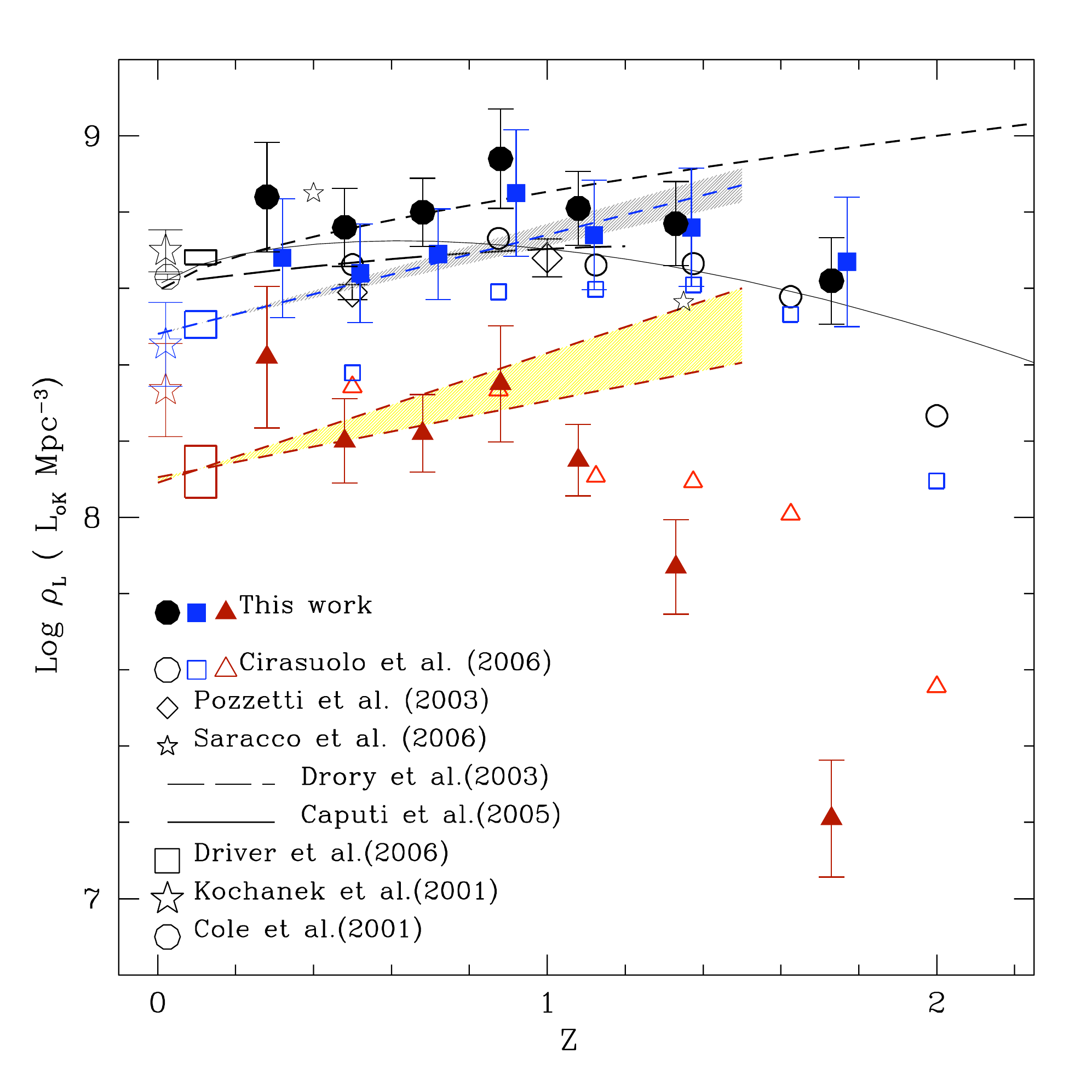}}
 \caption{Evolution of the luminosity density, $\rho_{L_K}$ (expressed
 in $L_{\odot K}$ ), for the total (filled circles) , active (filled squares)
 and quiescent (filled triangles)  populations. 
 Results are compared with previous works as specified in the caption and 
 with a pure luminosity evolution models (dashed lines).} 
 \label{fig:ld}
\end{figure}

 A sensitive test  for galaxy formation is the measurement of the 
 luminosity density which provides the total amount of light emitted
 per unit of volume and is estimated as: 
 $\rho_{L_K}=\int_0^{\infty}L\Phi(L)dL=\Phi^{\star}L^{\star}_K\Gamma(\alpha+2)$, 
 where $\Gamma$ is the gamma function and $(\Phi^{\star},L^{\star},\alpha)$ 
 are the Schechter parameters listed in Table~\ref{tab:lf}.  
 The measurements of $\rho_{L_K}$  (expressed in solar unit with $M_{K\odot}^{AB}=5.14$) are listed in Table~\ref{tab:lf} and  plotted in  Figure~\ref{fig:ld}. The quoted errorbars
 include all the sources of errors discussed in previous sections. \\
%
%
 The total sample shows a luminosity density increase 
 by a factor $\sim 2$  (starting from local values) up  to  $z\sim 1-1.2$ 
 followed by a fall off of a similar amount  up  to $z=2$. 
 This global behavior agrees, within the errorbars, with previous K band 
 luminosity density studies and is well described by the fit proposed by 
 Caputi et al. (2005, solid line).
 Looking at the SED selected samples, the active sample dominates the K band
 luminosity density at all z and follows a similar evolution as the global sample.
 For the quiescent sample, the luminosity density remains roughly flat up
 to $z\sim 1.2$ then drops significantly.   As discussed in previous section, this trend
  is less pronounced  in the UKIDSS 's red sample measurements. 
  
 To interpret the evolution of the luminosity density for the different samples,
 we need to set local references.
 Since no similar K-band based selection exists,
 we make use of existing optical catalogs with information
 that can be converted to yield a K band local reference. 
 We have used the results reported by  Driver et al. (2006) based on
 the Millennium Galaxy Catalog (MGC) where they provide the LFs in 
 B band and stellar mass density ($\rho_{\star}$) for a large variety of 
 classification, based on morphology, colors, SED fitting and spectral classes
 ($\eta$  parameter from 2dF:  Madgwick et al.; 2002).
 We adopt two criteria for the local reference  that can reflect the
 definition of our quiescent sample:
 \begin{itemize}
 \item As noted by Madgwick et al. (2003), the  $\eta$  parameter is highly
  correlated with the birthrate  b parameter. In particular their first class
  with $\eta \le -1.4$ corresponds to $b\le 0.1$ and is  thus similar to 
  what we expect locally for our criterion. We adopt
  the $\eta_1$ class as representative of our quiescent sample and define 
  all other classes as belonging to the active sample
  as reported in Table 2 of Driver et al. (2006)
 \item The majority of galaxies in the $(NUV-r')$ red sequence observed
  by GALEX  are dominated by de Vaucouleur profiles (Salim et al, 2005).
  We adopt the E/SO(red) morphological class of the MGC as reported in
  their Table 2 as another representation of our quiescent sample.
 \end{itemize}    

 Finally, we converted $\rho_{\star}$ to $\rho_{L_K}$ for each sample by
 adopting the  Bell and de Jong (2001) relations between (B-R) and M/L$_K$.
 The resulting ranges of local estimates are shown in the Fig~\ref{fig:ld} as colored 
 rectangles. We additionally plot the morphologically separated luminosity
 densities from 2MASS (Kochanek et al. , 2001) as colored open stars.
 The low-z estimates agree reasonably well betwen themselves except for
 the quiescent sample from 2MASS. 
 This is most likely due to the inclusion of the blue compact ellipticals in
 the sample of Kochanek et al. (2001). The MGC with
 its red E/S0 sample  appears to be most suited to provide our local
 reference in this case.
  
 Using the local normalization, we can now compare the evolution of 
 $\rho_{L_K}$ for pure luminosity evolution (PLE) models, i.e. with  
 no merging involved. We normalize to the local density $\Phi^{\star}(z=0)$ 
 and use $\Delta Log \rho_{L_K}(z)= -0.4 \Delta K_{ABS}(z)$. 
 We adopt the luminosity evolution as derived in Section \ref{sec:lf}. 
 Specifically, for the total sample we adopt the luminosity evolution from 
 Caputi et al. (2005), for the active sample we 
 adopt  $\Delta K_{ABS}(z)=-0.65(\pm0.07) \times Z$, while for the quiescent 
 sample we use two extrem PEGASE models that encircles the observed 
 luminosity evolution  and can be described by $\Delta K_{ABS}(z)= -0.5 | -0.85 \times Z$.
 In Fig~\ref{fig:ld}, the PLE predictions are shown with dashed lines
 and shaded areas.
 The evolution of the rest-frame K-band  luminosity density is consistent with the 
 PLE model for the total and active samples up to $z=1.2$, while 
 the model overpredicts $\rho_{L_K}$ at  higher redshift. This 
 shows that the number density of galaxies  has to drop. 
 For the quiescent sample, the PLE model fails to reproduce 
 the observed trend at low and high redshifts. 
 It predicts an increase in luminosity density by a factor 1.6 to 2.2, 
 up to $z\sim1.2$, while the observations suggest a  modest increase by a 
 factor ranging from 1.0 to 1.4. The disagreement with the PLE model at 
 highest redshift is even more pronounced than for the active sample 
 suggesting that the  number density of the quiescent galaxies must drop
 even faster.

%
\section{The stellar mass density up to z=2 } 
\begin{figure*}
\includegraphics[width=14cm]{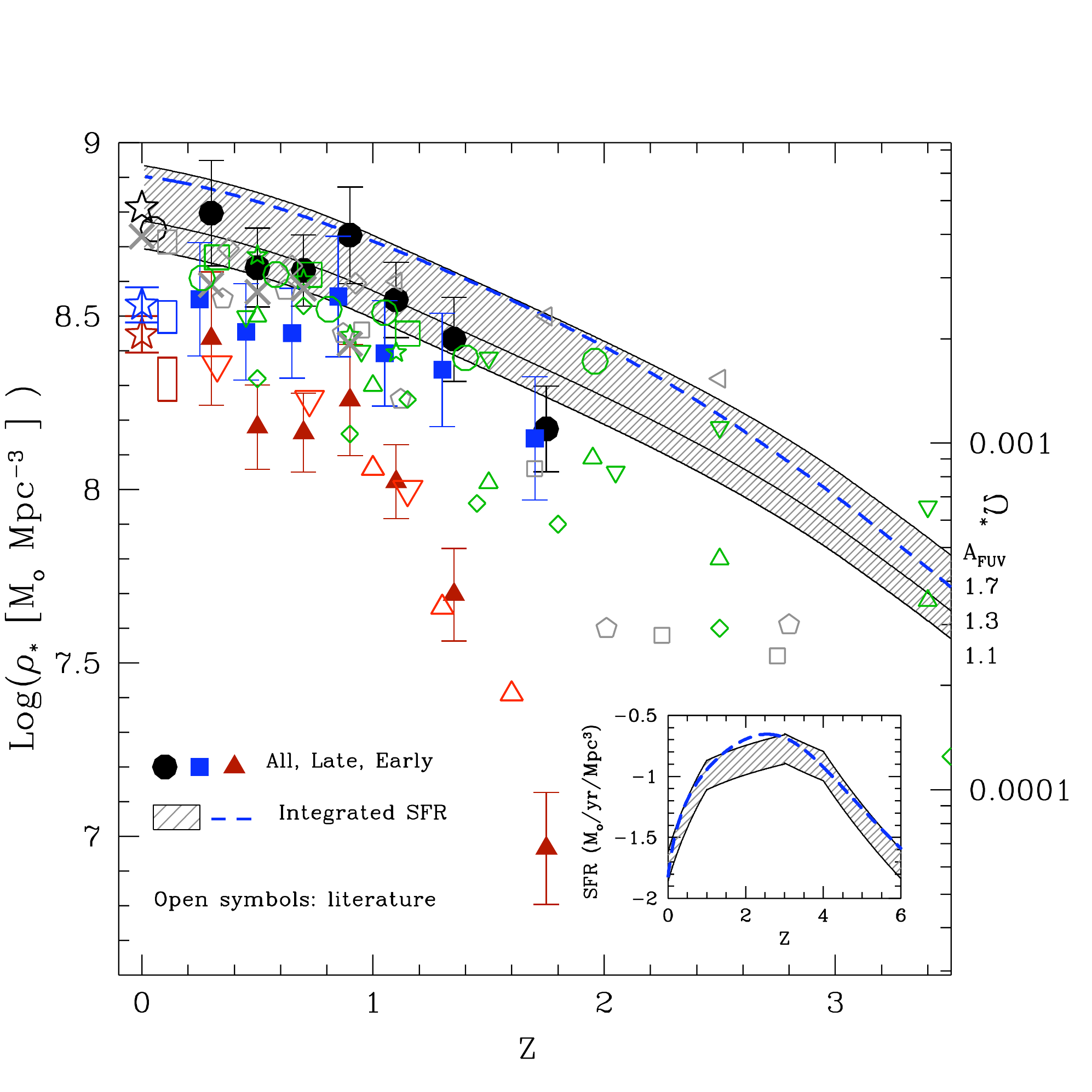}
 \caption{Evolution of the stellar mass density as a function of cosmic time
 (assuming a Salpeter IMF). The  total, active and quiescent stellar mass densities
 from this work  are shown with large filled circles, blue squares and red
 triangles respectively. For reference,
 the right-hand axis gives the stellar density parameter.
 The integrated Star Formation Rates for different dust attenuation corrections
 (A$_{FUV}$ = 1.1,1.3,1.7), based on the SFR derived by GALEX (Schiminovich et al., 2005),
 are plotted as solid lines, and the one from the compilation 
 of Hopkins and Beacom as a dashed line. The dust corrected SFRs are shown
 in the inset. High z measurements from the  literature
 for total samples have been  splitted between analysis based on optical information only
  (grey symbols) and including Near or Mid IR data  (green symbols) for the mass estimates. 
 with Optical : Brinchmann \& Ellis (2000; pentagons);
 Cohen et al. (2002;  losanges); Dickinson et al. (2003; squares);
 Gwyn et al. (2005;  inclined triangles);   Borch et al. (2006; crosses). 
 with NIR data: Drory et al. (2004; open stars) ; Drory et al. (2005; up and down   
 triangles);  Fontana et al. (2006; losanges);  Franceschini et al. (2006;  ellipticals: red down triangles and global samples : open squares);  Pozzetti et al. (2007; open circles); 
 Abraham et al. (2007; ellipticals : red up triangles). 
 Local values are from  Kochanek et al. (2001) for morphologically 
 selected ellipticals (red star) and spirals (blue star) and whole sample (black star);
 Cole et al. (2001; circle) and  Driver et al. (2006; colored  rectangles).
 For clarity we do not show errorbars for other surveys.
 }
\label{fig:rho}
\end{figure*}

 We now derive the stellar mass density, $\rho_{\star}$,  up to z=2.
 To that end we convert our K-band luminosity densities  (Table~\ref{tab:lf}), 
 to stellar mass  densities via the  relation
  $\rho_{\star}(z)=\rho_{L_K}(z) \times <\mbox{M/L}_K>(z)$ and 
  using the mass to light ratio equations determined in Section \ref{sec:ml}.\\
 Our measurements of  $\rho_{\star}$ for the three samples are
 shown  in Figure~\ref{fig:rho}. The errorbars  account for Poisson,  
 photo-z and cosmic variance  uncertainties (as for $\rho_{L}$)
  and  an additional uncertainty of 0.05dex from M/L estimates 
 (assuming that the paramaters a and b in the M/L relation are un-correlated). 
%
%
\subsection{The evolution of the stellar mass density}
 When moving back in time, the global population shows a
 small but regular decline up to $z\sim 1.1$ which accelerates
 at higher z. By comparing with local estimates,  
 we find that the total stellar mass has decreased by roughly a factor 
 of $\sim$ 1.5, 2 and 4  up to $z\sim 1.0, 1.5, 2.0$. 
 The compilation of  previous surveys shows a large scatter by 
 roughly a factor two and  our estimates are  located in the 
 upper envelope.  While the scatter can be in part due to mass estimates,
  for example the use or not of near IR data, the cosmic variance is most
 likely to be the dominant factor as discussed here and by Bell et al. (2003).\\
%
%
%
 We quantify the stellar mass evolution of the active and 
 quiescent samples  with a simple linear fit with redshift up to $z=1.2$.
   Including local measurements,  we get : 
   $\Delta Log\rho_{\star}^{Active}=-0.05 (\pm 0.09) z + 8.51 (\pm 0.04)$ and 
   $\Delta Log\rho_{\star}^{Quiescent}=-0.31 (\pm 0.07) z + 8.38 (\pm 0.02)$. 
 The active population shows a  modest evolution, consistent with no evolution
 with a mean value $\rho_{\star}^{Active}=10^{8.49\pm0.04}$. 
 This constancy of the active (blue) sequence has been pointed out by 
 Borch et al. (2006)  and by Martin et al. (2007) who derived  a similar value.
 On the other hand, we observe that the stellar mass of the quiescent galaxies,
 $\rho_{\star}^{Quiescent}$,  has increased  by a factor  2$\pm0.3$  between 
  $z= 1.2$ and $z=0$.
  Between $z=2$ and $z= 1.2$ the evolution in stellar mass is even stronger, 
  it increases by a factor of $\sim$10. This suggests two different regimes 
  in the build-up of the quiescent population. \\ 
%
%
\subsection{Comparison to the integrated star formation rate estimates}
 Thanks to the recent constraints on the cosmic star formation rate, 
 we can compare  the observed stellar mass density  ($\rho_{\star}$) 
 to the predictions for the cosmic star formation rate
 ($\dot{\rho}_{\star}$) over time, since the two quantities are simply related 
 as follows:  
 \begin{equation} 
 \rho_{\star}(t) = (1-R) \int_{0}^t \dot{\rho}_{\star}(t) dt
 \end{equation}
 where R is the recycling factor (fraction of stellar mass 
 released in the interstellar medium) which is assumed, in the
 case of a  Salpeter IMF, to  be $R=0.3$ (Madau et al.; 1998). 
 For the SFR, $\dot{\rho}_{\star}(z)$,  we use the uncorrected UV
 SFR as measured by Schiminovich et al. (2005) between $0\le z\le 3$
 that we have extended up to $z\sim 6$,  based on measurements
 from Steidel et al. (1999),  Bowens et al. (2006) and  Bunker et al. (2004).
 To derive the dust corrected UV-SFR, we adopt a range of 
 dust attenuation that varies between $1.1\le A_{FUV} \le 1.7$ which is 
 consistent with local estimates (Buat et al., 2005)  and with 
 other dust-free SFR estimators up to $z\sim1$  (Schiminovich et al., 2005;
 Hopkins and Beacom, 2006).   
 We also use the SFR estimate from the analysis of Hopkins and Beacom (2006) 
 who compile most of the recent surveys from the UV to the Far-Infrared
 to derive a global, dust corrected SFR \footnote{ We adopt their 
 SFR based on the modified Salpeter IMF which is converted 
 to a Salpeter IMF by applying +0.2dex in stellar mass.}. The two SFR histories
 are shown in an inset in Fig~\ref{fig:rho} and are in good agreement
 within the range of dust attenuations we have considered.
 The result of integrating $\dot{\rho}_{\star}$ is  shown as a
 shaded area. We emphasize that we did not impose beforehand that our
 integrated SFR  be consistent with the stellar mass density at $z=0$. 
 Nevertheless, it provides a remarkable fit in the redshift range probed in this work 
 and encloses well the local estimates  (see also Rudnick et al., 2006).

 We find that our measurements are well represented by  an
 attenuation value ranging from $1.1\le A_{FUV} \le 1.3$) between $0\le z\le 1.5$.
 This constancy of the mean dust attenuation up to $z\sim 1.5$,
 can be  qualitatively interpreted as a trade-off between
 two competing factors:  -1- star-forming galaxies at high-z have less dust
 attenuation than their local counterparts at the same total luminosity
 (Reddy et al., 2006, Xu et al., 2007) -2-  star-forming galaxies at high-z
   are intrinsically more luminuous (Arnouts et al., 2005;   Le Floch et al., 2006). 
 We can quantify these effects by looking at the relations between
 the ($L_{Dust}/L_{UV}$) ratio and the total luminosity 
 ($L_{TOT}=L_{Dust}+L_{UV}$), all quantities expressed as $\nu L_{\nu}$ .
 From the local relation (Martin et al. , 2005),
 a typical $L^{\star}_{UV}(z=0)$ galaxy has a  $L_{Dust}/L_{UV}\sim 3.75$, 
 corresponding to $A_{FUV}\sim1.3$ (Buat et al., 2005).
 If we let this galaxy evolve according to the evolution of the luminosity functions 
 ($L_{FUV}\sim (1+z)^{2.5}$;  Arnouts et al., 2005 and  $L_{FIR}\sim (1+z)^{3.2}$;
 Le Floch et al., 2006), at $z\sim 1.35$  we obtain a typical luminosity of 
 $log(L_{TOT}/L_{\odot})\sim 11.34$. 
 We can finally estimate  the amount of dust attenuation for this galaxy based on the 
 relation from Reddy et al. (2006, their Eq. 5)
 that holds in the redshift range $1\le z\le 3$. 
 We obtain  $L_{Dust}/L_{UV}\sim 4.3$, corresponding to $A_{FUV}\sim1.4$.
 This rough estimate shows that a typical $L^{\star}(z)$ galaxy that
 contributes to the bulk of the SFR at all z shows, on average, a small evolution
 of the amount of dust attenuation, in agreement with the observation of the integrated 
 dust corrected SFR in Fig~\ref{fig:rho}.

 To summarize, with  simplified but not unrealistic estimates of the dust correction, 
 we observe that the two complementary quantities, the star formation
 rate and the  stellar mass density, are remarkably consistent with each other.
 This result is encouraging, and may provide some more fundamental 
 informations about the initial mass function (IMF).   
 The observations of the two quantities explore two different parts of
 the IMF (dominated by massive stars, a few $M_{\odot}$, for the SFR, 
 and low mass stars, for stellar mass) and are extrapolated through
 the adopted IMF.  The good agreement of these complementary observations
 over the last 10 Gyrs, is a good support for  an IMF that is on average universal,
 as already pointed out in previous work (e.g. Franceschini et al., 2006). 
 We note that this remark is not specific to the Salpeter IMF  adopted throughout 
 this paper, but holds also for other IMFs suggested in the literature, as long as 
 the quantities ($\rho_{\star}$ , $\dot{\rho}_{\star}$) are appropriately corrected 
 for IMF effects.
%
%
%
\begin{table*}
\centering
\caption[]{ K-LF parameters ($\alpha$, $M^{\star}$ , $\Phi^{\star}$) and K luminosity densities ($\rho_L$) for the whole, active and quiescent samples.
 Errors in $M^{\star}$ , $\Phi^{\star}$ refer to Poisson errors and slope uncertainties. 
 Additional uncertainties due to photometric redshift (PdZ) and cosmic variance (CV)
are reported separately. Errors in  $\rho_L$ include all of them.
\label{tab:lf}}
\begin{tabular}{|cccccccc|}
\hline
 $<z>$  & \# & $\alpha$   & $M^{\star}$    &  $\Phi^{\star}$            &  $(\frac{d\Phi}{\Phi})_{PdZ}$  & $(\frac{d\Phi}{\Phi})_{CV}$  & $\rho_L$ \\
            &      &                 &                        &(10$^{-3}$ Mpc$^{-3}$)&              &                      &(10$^8 L_{\odot K}$Mpc$^{-3}$)\\
\hline
       &         &                   &    ALL              &                                  &               &                     &             \\
\hline
0.3  & 2180 & $-1.1 \pm 0.2$  & $-22.84_{-0.6}^{+0.4}$ & $4.19_{-1.9}^{+2.2}$ & 0.05 & 0.23  &   $6.97\pm2.2$  \\
0.5  & 2680 & $-1.1 \pm 0.2$  & $-22.83 \pm 0.3$     & $3.50 \pm 1.3$                 & 0.05 & 0.20 &   $5.77\pm1.3$ \\
0.7  & 3336 & $-1.1 \pm 0.2$  & $-22.96 \pm 0.2$   & $3.36 \pm 1.1$  &  0.05  & 0.16  &  $6.24\pm1.3$ \\
0.9  & 4545 & $-1.1 \pm 0.2$  & $-23.08 \pm 0.2$    & $4.22 \pm 1.2$  & 0.05 & 0.26   &  $8.75\pm2.5$ \\
1.1  & 3027 & $-1.1 \pm 0.2$  & $-23.22 \pm 0.2$      & $2.75 \pm 0.7$  & 0.11 & 0.13   &  $6.48\pm1.4$ \\
1.35& 3077 & $-1.1 \pm 0.2$  & $-23.18 \pm 0.17$    & $2.56 \pm 0.5$  & 0.09 & 0.12  &  $5.82\pm1.5$ \\
1.75& 2189 & $-1.1 \pm 0.2$  & $-23.42 \pm 0.14$    & $1.47 \pm 0.2$  & 0.07 & 0.10  &  $4.17\pm1.1$ \\
\hline
         &         &                      &        active                            &                           &               &                       \\
\hline
0.3    & 1666 & $-1.3 -\pm 0.2$ & $-22.83_{-0.56}^{+0.43}  $ & $2.38_{-1.3}^{+1.6}$   & 0.05 & 0.27 &  $4.76\pm1.7$  \\ 
0.5    & 2063 & $-1.3 \pm 0.2$  &  $-22.82 \pm 0.3$    & $2.22 \pm 1.0$     & 0.05  & 0.21 &  $4.40\pm  1.3$  \\
0.7    & 2569 & $-1.3 \pm 0.2$  &  $-22.95 \pm 0.2$    & $2.21 \pm 0.9$     & 0.05  & 0.17 &   $4.94\pm 1.3$ \\
0.9    & 3428 & $-1.3 \pm 0.2$  &  $-23.06 \pm 0.2$    & $2.82 \pm 1.0$     & 0.05  & 0.28 &  $6.97\pm  2.7$ \\
1.1    & 2358 & $-1.3 \pm 0.2$  &  $-23.19 \pm 0.2$    & $1.99 \pm 0.6$     & 0.12  & 0.14 &  $5.55\pm  1.8$  \\
1.35  & 2609 & $-1.3 \pm 0.2$  &  $-23.19 \pm 0.18$  & $2.08\pm 0.5$      & 0.09  & 0.13 &  $5.80\pm  2.0$ \\
1.75  & 2081 & $-1.3 \pm 0.2$  &  $-23.49 \pm 0.15$  & $1.27 \pm 0.3$     & 0.07  & 0.11 &  $4.66\pm  1.8$ \\
\hline 
         &         &                      & quiescent         &                           &               &                     \\
\hline 
0.3    &   514  & $-0.6\pm 0.2$ &$-22.91_{-0.49}^{+0.34}$& $1.78_{-0.5}^{+0.42}$& 0.09 & 0.33  &   $2.63 \pm 1.1$   \\
0.5    &   617  & $-0.3\pm 0.2$   & $-22.55 \pm 0.19$        &   $1.47 \pm 0.17$       & 0.07 & 0.24 &   $1.60  \pm 0.4$  \\
0.7    &   767  & $-0.3\pm 0.2$   & $-22.73 \pm 0.15$        &   $1.29 \pm 0.13$       & 0.09 & 0.21 &   $1.65 \pm 0.4$  \\
0.9    & 1117  & $-0.3\pm 0.2$   & $-22.83 \pm 0.14$        &   $1.58 \pm 0.14$       & 0.07 & 0.34 &   $2.22 \pm 0.8$  \\
1.1    &   669  & $0.0 \pm 0.3$   & $-22.86 \pm 0.17$        &   $0.90\pm 0.08$        & 0.12 & 0.17 &   $1.42 \pm 0.3$  \\
1.35   &  468  & $0.3 \pm 0.3$   & $-22.84 \pm 0.16$        &   $0.41 \pm 0.03$       & 0.21 &  0.18 &  $0.74 \pm 0.2$ \\
1.75  &   108  & $0.6 \pm 0.3$   & $-23.27 \pm 0.15 $       &   $0.05 \pm 0.02$       & 0.28 & 0.17  &   $0.17 \pm 0.05$ \\
\hline
\end{tabular}
\end{table*}
%
%
\section{Discussion and Conclusions}

 In this work, we have used a unique large sample
 of 21200 galaxies selected at 3.6$\mu m$ (based on the VVDS and SWIRE surveys) 
 to investigate the evolution of the luminosity functions, luminosity densities,
 stellar mass to light relations and stellar mass densities up to $z\sim2$.
 We have separated the active and quiescent galaxies based on an 
 SED  fitting procedure. We define as quiescent those galaxies which 
 are best fit by an elliptical template.
 We have shown that this sample reproduces well the red sequence of the
 color bimodality $(NUV-r')$ (known to be a good separator between active and   
 quiescent galaxies with $b\le 0.1$) and at the same time minimizes the
 contamination by dusty starbursts.
 We find that the active and quiescent populations follow different  behaviours. 
 In particular, there is a clear transition between two regimes in the evolution of the 
 quiescent population at a redshift of $z=1.2$ (i.e. 8 Gyr ago). 
 
\subsection{The Last 8 Gyrs}

 The active  population appears to be in place at $z\sim 1.2$
 with a small evolution over this time laps except the aging of its stellar population. In contrast the  
 quiescent sequence shows a gradual increase  with a doubling of its
 stellar mass in line with previous optical studies (Bell et al., 2003;  
 Faber et al., 2006; Brown et al. (2006).

 The lack of evolution of the stellar mass density for the active, $\rho_{\star}^{active}$,
 and the doubling for quiescent, $\rho_{\star}^{quiescent}$, is surprising. 
 The gradual increase of the global stellar mass density, over 
 the last 8 Gyrs can be easily explained by the formation of new stars by star 
 forming galaxies at a rate described by the cosmic star formation rate. Or,
 similarly, it is found that intermediate mass galaxies 
 ($M/M_{\odot}\sim 10.5$) have specific SFRs allowing them to
 increase their stellar mass by a factor two since $z=1$ (Juneau et al., 2005).
 On the other hand, the increase  of $\rho_{\star}^{quiescent}$  is in apparent
 contradiction with the definition of quiescent galaxies that cannot produce
 more than 5\% of additional stellar mass over the last 8Gyr (due to the 
 absence of star formation). The most plausible explanation for the increase
 of $\rho_{\star}^{quiescent}$ is a progressive migration of galaxies
 from the active to the quiescent population at a similar rate 
 than that with which new stars are formed. \\
 We quantify this evolution by a rough estimate of the stellar mass growth per unit
  of time  defined as :
 \begin{equation} 
 \dot{\rho}_X = \frac{\rho_{\star}^X(z_l)-\rho_{\star}^X(z_h)}{t_{univ}(z_l)-t_{univ}(z_h)}
 \end{equation}
 , where $X$ refers to stellar mass of the considered sample in  the redshift range
 $z_l\le z \le z_h$.  
 We find that  the stellar mass growth of the active population 
 evolves very little with  $\dot{\rho}_{Active}= 0.005\pm0.003 M_{\odot}/Mpc^3/yr$, 
 while the  quiescent population has a much higher stellar mass growth with
 $\dot{\rho}_{Quiescent}= 0.017\pm0.004 M_{\odot}/Mpc^3/yr$.  
  We can compare this value with the mass growth expected by integrating 
 the star formation history over the same period of time. 
 We find that $\dot{\rho}_{SFR}= 0.025\pm0.003 M_{\odot}/Mpc^3/yr$
 (assuming  $1.1\le A_{FUV}\le 1.3$).
 %
 Under the assumption that the quiescent population has negligible
  residual star formation, its mass growth can be attributed to the mass flux
  of active galaxies moving into a quiescent mode
  ($\dot{\rho}_{A\rightarrow Q}=\dot{\rho}_{Quiescent}$),
  and which appear to account for most of the global stellar mass growth 
  derived from  the SFR.\\
Additional evidence for this scenario is also presented by 
Vergani et al. (2007) who split the VVDS spectroscopic sample in
 active and quiescent galaxies based on  4000 {\AA} break. 
In order to satisfy the constraints from the stellar mass functions split
in active and quiescent galaxies, they find that the star formation in
some massive blue galaxies must have been quenched, moving these 
galaxies into the "red sequence".

%
\subsection{The major epoch of build-up for quiescent galaxies:}
 The present analysis allows us to extend previous work to the redshift range between 
 $2\ge  z \ge 1.2$  ($3.2Gyr \le T_{univ} \le 5Gyr$). This appears to be 
 a transition epoch, as evidenced by the strong increase in stellar mass, 
 by a factor of $\sim$10, of the quiescent population between $2\ge z\ge 1.2$, 
 while the active population increases still by a factor $\sim$2.5. 
 This translates into a stellar mass growth for the quiescent populations of 
 $\sim0.065 M_{\odot}/Mpc^3/yr$, which is more than 3 times faster than the 
 evolution at $z\le 1.2$.  In contrast to $z\le 1.2$, where most of the galaxies
 seem to be in place,  at  $z\ge 1.2$, galaxies are still in an active phase of
 mass assembly. In particular, the evolution of the quiescent population 
 suggests that we are observing the epoch when , for the first time in the history 
 of the universe, a large number of active galaxies
 are ending their star formation activity and start to build up a quiescent population.
 While the mechanism acting in this process is not clear, gas exhaustion, merging
 or other effects, this build-up happens a few Gyr after
 the peak of the cosmic SFR (Hopkins and Beacom, 2006). An interesting related
 information is the similar evolution followed by morphologically selected elliptical
 galaxies  (Franceschini et al., 2006; Abraham et al., 2007).   
 If not by chance, this coincidence could suggest that the 
 build-up of the quiescent sequence is closely 
 followed or preceded by  a  morphological  transformation.
\begin{acknowledgements}
 We are grateful to the referee for his careful reading of the manuscript and his suggestions.
 SA wants to thank B. Milliard for useful discussions.
This work has been partially supported by the CNRS-INSU and
the "Progamme National de Cosmologie" (France), 
the "Programme National Galaxies" (France) and
a grant awarded for "EFIGI project" (grant \# 45500) 
from the French research ministry.
CJW is supported by the MAGPOP Marie Curie EU Research and Training Network.

\end{acknowledgements}

\end{document}